\documentclass[useAMS,usenatbib,usegraphicx]{mn2e}

\usepackage{times}

\newcommand\hi{\mbox{H\,{\sc i}}}

\title[M31 Substructure Star Formation Histories]{The Nature and
  Origin of Substructure in the Outskirts of M31 -- II. Detailed Star
  Formation Histories\thanks{Based on observations made with the
    NASA/ESA Hubble Space Telescope, obtained at the Space Telescope
    Science Institute, which is operated by the Association of
    Universities for Research in Astronomy, Inc., under NASA contract
    NAS5-26555. These observations are associated with programmes
    GO-9458 and GO-10128.}}
\author[E.~J.\ Bernard et al.]{%
Edouard J. Bernard,$^{1}$\thanks{E-mail: ejb@roe.ac.uk}
Annette M. N. Ferguson,$^{1}$
Jenny C. Richardson,$^{1,2}$
\newauthor
Mike J. Irwin,$^{2}$
Michael K. Barker,$^{1}$
Sebastian L. Hidalgo,$^{3,4}$
Antonio Aparicio,$^{3,4}$
\newauthor
Scott C. Chapman,$^{2,5}$
Rodrigo A. Ibata,$^{6}$
Geraint F. Lewis,$^{7}$
\newauthor
Alan W. McConnachie,$^{8}$
Nial R. Tanvir$^{9}$ \\
 $^{1}$SUPA, Institute for Astronomy, University of Edinburgh, Royal
   Observatory, Blackford Hill, Edinburgh EH9 3HJ, UK \\
 $^{2}$Institute of Astronomy, Madingley Road, Cambridge, CB3 0HA, UK \\
 $^{3}$Instituto de Astrof\'{i}sica de Canarias, Calle V\'ia L\'actea s/n,
   38205 La Laguna, Tenerife, Spain \\
 $^{4}$Departamento de Astrof\'{i}sica, Universidad de La Laguna, 38200
   Tenerife, Spain \\
 $^{5}$Department of Physics and Atmospheric Science, Dalhousie University,
   6310 Coburg Road, Halifax, Nova Scotia B3H 4R2, Canada \\
 $^{6}$Observatoire Astronomique de Strasbourg, Universit\'e de Strasbourg,
   CNRS, UMR 7550, 11 rue de l'Universit\'e, F-67000 Strasbourg, France \\
 $^{7}$Institute of Astronomy, School of Physics, University of Sydney,
   NSW 2006, Australia \\
 $^{8}$NRC Herzberg Institute for Astrophysics, 5071 West Saanich
   Road, Victoria, V9E 2E7, British Columbia, Canada \\
 $^{9}$Department of Physics \& Astronomy, University of Leicester,
   Leicester LE1 7RH, UK
}

\begin{document}

\date{Accepted --. Received --; in original form --}

\pagerange{\pageref{firstpage}--\pageref{lastpage}} \pubyear{2014}

\maketitle

\label{firstpage}

\begin{abstract}
 While wide-field surveys of M31 have revealed much substructure at
 large radii, understanding the nature and origin of this material
 is not straightforward from morphology alone. Using deep HST/ACS
 data, we have derived further constraints in the form of quantitative
 star formation histories (SFHs) for 14 inner halo fields which sample
 diverse substructures. In agreement with our previous analysis of
 colour-magnitude diagram morphologies, we find the resultant behaviours
 can be broadly separated into two categories. The SFHs of `disc-like'
 fields indicate that most of their mass has formed since $z\sim1$, with
 one quarter of the mass formed in the last 5~Gyr. We find `stream-like'
 fields to be on average 1.5~Gyr older, with $\la$10 percent of
 their stellar mass formed within the last 5~Gyr. These fields
 are also characterised by an age--metallicity relation showing rapid
 chemical enrichment to solar metallicity by $z=1$, suggestive of an
 early-type progenitor. We confirm a significant burst of star formation
 2~Gyr ago, discovered in our previous work, in all the fields studied
 here. The presence of these young stars in our most remote fields
 suggests that they have not formed {\it in situ} but have been
 kicked-out from the thin disc through disc heating in the recent past.
\end{abstract}


%-------------------------------------------------------------------------------

\section{Introduction}

\begin{figure}
\includegraphics[width=8.8cm]{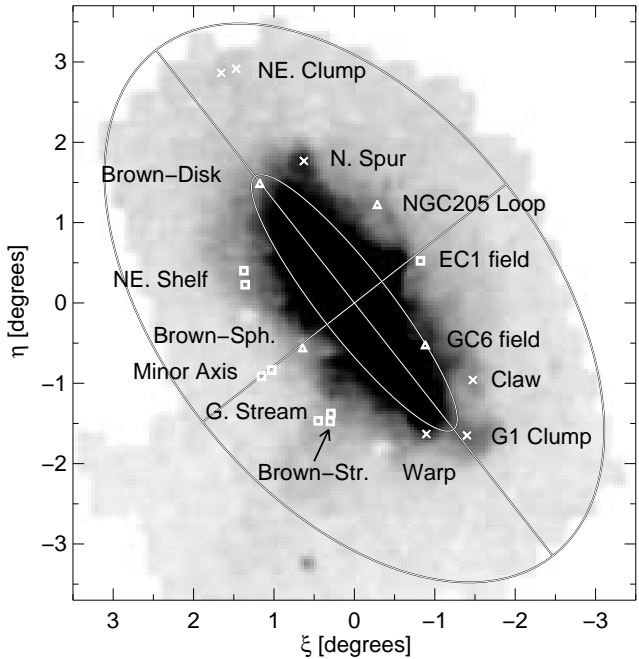}
\caption{Locations of our {\it HST/ACS} pointings superimposed onto
  the INT/WFC map of M31's inner halo \citep{irw05}, showing the
  distribution of evolved giant stars around M31. Crosses, open
  squares, and open triangles represent the disc-like, stream-like,
  and composite fields (see text for details).  The image spans 95~kpc
  $\times$ 100~kpc and each ACS field covers $0.8~{\rm kpc} \times
  0.8~{\rm kpc}$ at the distance of M31.  All of the significant
  substructure discovered during the course of the INT/WFC survey is
  sampled by our data.  The inner ellipse has a semimajor axis of
  2$\degr$ (27~kpc) and represents an inclined disc with $i$ = 77\fdg
  5 and position angle of 38\fdg 1.  The outer ellipse, of semimajor
  axis length 4$\degr$ (55~kpc), roughly indicates the spatial extent
  of the INT survey.}\label{fig:map}
\end{figure}

Galaxy evolution proceeds dynamically through interactions, accretions
and mergers with other dark matter halos and through the formation and
evolution of the stars within them. Such events may influence the
local star formation rate and/or add new material directly from the
tidal disruption of satellite galaxies.  The fossil record of a
galaxy's history is encoded within the detailed properties of its
stellar populations. Over the last decade, a variety of methods have
been applied to mine this information in both the local and distant
Universe.

Retrieving the star formation history (SFH) from a deep
colour-magnitude diagram (CMD) is one of the most powerful tools
available for understanding galaxy assembly.  Since stars occupy very
specific loci on the CMD depending on their evolutionary phases, the
morphology of a CMD can be interpreted in terms of the underlying
composition of the parent population. By comparing an observed CMD
with a large library of synthetic CMDs constructed from theoretical
stellar evolution models and various input parameters (e.g.\ initial
mass function [IMF], binary fractions), the temporal evolution of the
star formation rate and chemical enrichment history can be extracted.

Quantitative star formation history work is, to some extent, still in
its infancy. While a great deal of progress has been made in charting
the SFHs of dwarf galaxies in the Local Group \citep[LG; see][and
references therein]{tol09,wei14}, these constitute only a small fraction
of the local stellar mass density.  In the Milky Way, our knowledge of
the detailed SFH is essentially confined to a $\la 100$~pc region in
the disc around the Sun, and even then with large uncertainties at
both young ($\la 1$~Gyr) and old ($\ga 7$~Gyr) ages
\citep[e.g.][]{cign06}.  The SFH across the M33 disc is reasonably
well known \citep{wil09, bar11} but a much less detailed picture
exists for our sister galaxy, M31.  This is due in part to the large
angular size of M31 on the sky and the fact that the inner
$\sim$20~kpc has a high stellar density and significant differential
reddening \citep[e.g.][]{bed10,dal12}, hindering the ability to
reliably probe the ancient SFH of the main body.

Resolving the oldest main-sequence turn-off (MSTO) stars in M31 --
crucial for completely breaking the age--metallicity degeneracy --
requires many orbits of the {\it HST} \citep[e.g.][]{bro03}; yet a
single Advanced Camera for Surveys {\it ACS} field ($0.8 \times
0.8~{\rm kpc}$ at the distance of M31) captures but a tiny fraction of
the galaxy's extent.
One of the first attempts to extract the detailed SFH in M31 back to
early epochs was carried out by \cite{bro06}. These authors obtained
ultra-deep {\it HST/ACS} imaging of three fields (nominally in the outer
disc, giant stellar stream [GSS] and halo) down to the oldest MSTO. They
found evidence for extended star formation and strong intermediate-age
components (ages $\sim 4-8$~Gyr), with slight variations from field to
field. Unfortunately, interpreting these results in the context of the
M31's formation history was not straightforward given the rich
substructure present in all the fields.  More recently, we have
explored the SFH of a field located in the south-western warp of the
main disc \citep{ber12}. Our derived SFH is far from smooth and
includes the presence of a strong burst of star formation about 2~Gyr
ago that lasted about 1~Gyr and contributed $\sim$25~percent of the
total mass of stars formed in this field.  Given the M33 outer disc
exhibits a concurrent burst of star formation \citep{bar11, ber12}, we
suggested that these bursts might have been triggered by an
interaction between the two galaxies. Indeed, self-consistent N-body
modeling of the M31-M33 system suggests that their last perigalactic
passage occurred at about this epoch \citep{mcc09}.

%%%%%%%%%%%%%%%%%%%%%%%%%%%%%%%% TAB 1 %%%%%%%%%%%%%%%%%%%%%%%%%%%%%%%%%%%%%%%%%
\begin{table*}
\centering
 \begin{minipage}{107mm}
\caption{Field Information.}
\label{tab:obs} 
\begin{tabular}{ @{}ccccccc}
  \hline 
  Field & Type\footnote{S: stream-like; D: disc-like; C: composite.} & R.A. & Dec. & E(B $-$ V)\footnote{Values from \citet{sch98}.} & R$_{proj}$\footnote{Projected radial distance.}
  & R$_{disc}$\footnote{Radii within the disc plane calculated assuming an inclined disc with $i = 77.5\degr$ and a position angle of 38.1\degr.} \\
  name &  & (J2000) & (J2000) &  & (kpc) & (kpc) \\
  \hline  
  G Stream\footnote{Low stellar density fields that required two adjacent pointings.}       & S & 00:44:15.5 &  39:53:30.0 & 0.058 & 19.0 & 79.5 \\
                 &   & 00:45:05.0 &  39:48:00.0 & 0.051 & 20.7 & 68.4 \\
  Brown-stream   & S & 00:44:18.0 &  39:47:36.0 & 0.053 & 20.3 & 72.7 \\
  Minor Axis$^e$     & S & 00:48:08.4 &  40:25:30.0 & 0.060 & 17.9 & 91.8 \\
                 &   & 00:48:47.8 &  40:20:44.6 & 0.059 & 19.9 & 82.7 \\
  EC1\_field     & S & 00:38:19.5 &  41:47:15.4 & 0.070 & 13.2 & 60.5 \\
  NE Shelf$^e$       & S & 00:49:59.4 &  41:28:55.5 & 0.065 & 18.6 & 54.3 \\
                 &   & 00:50:05.7 &  41:39:21.4 & 0.071 & 19.3 & 59.5 \\
  \hline 
  G1 Clump       & D & 00:35:28.0 &  39:36:19.1 & 0.063 & 29.2 & 29.7 \\
  Warp           & D & 00:38:05.1 &  39:37:54.9 & 0.054 & 25.1 & 31.1 \\
  Claw           & D & 00:35:00.3 &  40:17:37.3 & 0.060 & 23.7 & 42.0 \\
  N Spur         & D & 00:46:10.0 &  43:02:00.0 & 0.079 & 25.3 & 43.3 \\
  NE Clump$^e$       & D & 00:51:56.7 &  44:06:38.5 & 0.093 & 44.1 & 59.0 \\
                 &   & 00:50:55.2 &  44:10:00.4 & 0.092 & 44.6 & 53.0 \\
  \hline 
  GC6\_field     & C & 00:38:04.6 &  40:44:39.8 & 0.074 & 13.8 & 26.1 \\
  NGC205 Loop    & C & 00:41:11.6 &  42:29:43.1 & 0.076 & 17.0 & 71.5 \\
  Brown-disk     & C & 00:49:08.5 &  42:44:57.0 & 0.080 & 25.6 & 25.6 \\
  Brown-spheroid & C & 00:46:08.1 &  40:42:36.4 & 0.081 & 11.5 & 53.0 \\
  \hline\vspace{-10mm}
\end{tabular}
\end{minipage}
\end{table*}

%%%%%%%%%%%%%%%%%%%%%%%%%%%%%%%%%%%%%%%%%%%%%%%%%%%%%%%%%%%%%%%%%%%%%%%%%%%%%%%%

The highly structured nature of M31's outer regions has been revealed
by a number of wide-field imaging surveys over the last decade
\citep[e.g.][]{iba01, fer02, iba07, tan10, iba14}.  Various features
are observed, such as streams, loops and shells, however it is
difficult to constrain the origin of this material on the basis of its
morphology alone.  While such features are often associated with
accretions of small dwarf galaxies, it is also possible that much of
the material is simply disrupted and heated disc that has been kicked
out during a recent encounter \citep[e.g.][]{kaz08}.  In attempt to
decipher the nature and origin of the brightest substructures in M31,
we have undertaken a deep {\it HST/ACS} survey of 14 fields in the
inner halo and outer disc, many of which lie on discrete stellar
substructures.  Our fields span projected galactocentric radii of
$12$~kpc $\la$ R$_{proj} \la 45$~kpc and are deep enough to resolve
individual stars down to three magnitudes below the horizontal-branch
(HB).  \cite{ric08} presented a comparative analysis of the CMDs of
these fields which led to the conclusion that the substructure seen
had two distinct origins. Five of the fields have stellar populations
identical to those of the GSS, the progenitor of
which is a putative $\sim10^9-10^{10}~{\rm M}_{\sun}$ galaxy that was
cannibalised by M31 some time in the recent past \citep{fard07}.
Another five fields show evidence for moderately young populations,
consistent with them having once been part of the thin disc of M31 and
subsequently heated by a tidal ancounter. The remaining four fields
show evidence of both stream-like and disc-like stellar populations
mixed together, which \citet{ric08} termed as composite fields.

Here we extend the analysis of \cite{ric08} by using the technique of
synthetic CMD fitting to provide a more detailed and robust
quantitative analysis of these fields. The method is fully consistent
with that presented in \citet{ber12}; the drawback is that our
photometry is slightly shallower ($\sim$1~mag) and therefore does not
capture the oldest MSTOs with good signal-to-noise. This limits the
information available about the earliest epochs of star formation and
tends to blur the age--metallicity relation (AMR). The paper is
structured as follows: in Section~\ref{obs}, we present the
observations and the data reduction steps. The CMD-fitting method is
described in Section~\ref{sfh}, along with the resulting SFHs. Our
interpretation of the results is discussed in Section~\ref{interp},
and a summary of the main results is given in Section~\ref{cl}.

%-------------------------------------------------------------------------------

\section{Observations and Data Reduction}\label{obs}

\subsection{The Dataset}\label{data}

The dataset consists of 14 deep HST/ACS fields in the inner halo and
outer disc of M31 and it has been fully described in \citet{ric08}.
Nine of the fields were chosen to specifically sample stellar
substructures in the inner halo of M31 (proposals GO-9458 and
GO-10128; P.I.: A.\ Ferguson). Another two fields were observed as
part of our imaging survey of halo globular clusters (GO-10394; P.I.:
N.\ Tanvir); to study the background M31 field populations, we have
masked the clusters by removing all the stars within their respective
tidal radii \citep{bar07,tan12}. The last three fields are archival
data from ultra-deep imaging programs (GO-9453 and GO-10265; P.I.: T.\
Brown) targeting the outer disc, spheroid, and the GSS.

\begin{figure*}
\includegraphics[width=17cm]{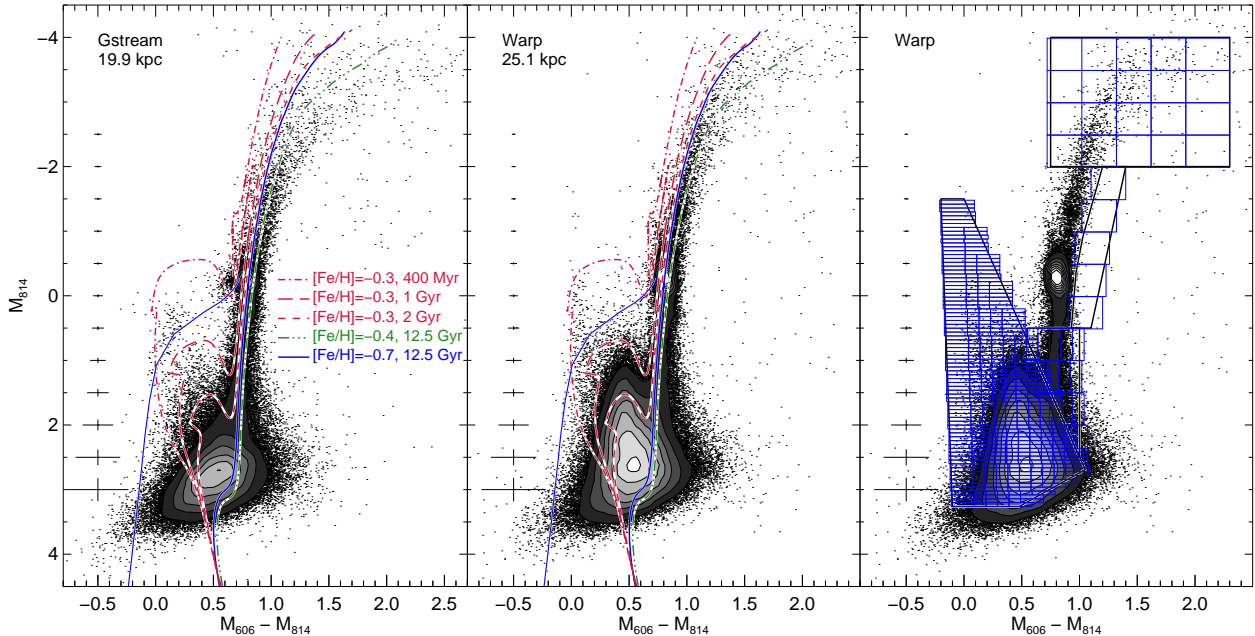}
\caption{Colour-magnitude diagrams for the {\it Giant Stream} (left)
  and the {\it Warp} (middle) fields, where selected isochrones and a
  ZAHB from the BaSTI library \citep{pie04} are overlaid. The error
  bars show the mean photometric errors as a function of magnitude.
  The projected radial distances are indicated in each panel. The
  contour levels correspond to [15, 25, 35, 45, 55, 65, 75]$\times
  10^3 {\rm \ stars \ mag}^{-2}$. The black and blue lines in
  the right panel show the location of the regions (`bundles') and
  the boxes, respectively, used for the CMD-fitting (see
  Section~\ref{method}), overplotted on the {\it Warp} CMD.}
\label{fig:cmd}
\end{figure*}

Figure~\ref{fig:map} shows the location of our fields superimposed
onto the INT/WFC map of the outer disc and inner halo of M31.  Most
targets were observed via a single pointing but the low stellar
density in four fields required two adjacent pointings to be obtained.
While the data analysed in this work typically comprises three {\it
HST} orbits per pointing, some of the fields were originally
observed for considerably longer. As explained in \cite{ric08}, we
have analysed only a subset of these data in order to match the depth
of our other {\it HST/ACS} fields and allow a homogeneous comparison.
Table~\ref{tab:obs} contains information about the location of the
fields, their distance from the centre of M31, and their colour excess
from \cite{sch98}.  We also indicate whether the fields were
identified as disc-like (D), stream-like (S) or composite (C) on the
basis of CMD morphology by \citet{ric08}. The substructures probed by
our fields are briefly described below.

\begin{itemize}
\item The \emph{NE Clump}\footnote{From here on we use italicised font
  for the name of our fields, to distinguish them from the name of
  the substructure they probe.}  probes a large ($1\degr$ $\approx$
  13~kpc in diameter) diffuse over-density of stars over 40~kpc away
  from the centre of M31 along the north-eastern major axis. This
  clump of stars appears to be connected to M31 by a faint filament.
  Using shallow SDSS imaging reaching the tip of the red giant branch (RGB),
  \cite{zuc04} noted a similarity between the stellar populations of
  the NE Clump and the G1 Clump (see below). They suggest the two may
  have been torn off from the thin disc or part of an ancient tidal
  stream, or that the NE Clump is a satellite galaxy of M31 undergoing
  tidal disruption.  \citet{kni14} recently suggested a possible
  association with the GSS on the basis of planetary nebula
  kinematics.
\item The \emph{NGC205 Loop} field probes an arc of material in the
  inner halo of M31 which appears to emanate from the dwarf elliptical
  satellite, NGC\,205.  The RGB colour and kinematics suggest that
  this loop could be material tidally stripped from NGC\,205
  \citep{mcc04,mcc05b, iba05}.
\item The \emph{Claw}, named for its peculiar morphology, is a
  significant substructure which protrudes from south-western part of
  the disc.
\item The \emph{G1 Clump} samples the prominent over-density of stars
  ($\sim$12~kpc in diameter) on the south-western major axis.
  Originally named for its proximity to the nearby massive globular
  cluster G1, subsequent stellar population analyses have since
  discounted a link between the two \citep{ric04, fari07}. Meanwhile,
  Keck spectroscopy has shown that the kinematics of the G1 Clump
  stars have more in common with the neutral hydrogen disc of M31 than
  with either the G1 globular cluster or the stellar halo
  \citep{rei04, iba05}.
\item The stellar and gaseous discs of M31 are warped at large radii
  \citep[e.g.,][]{inn82,bri84,wal88}; the \emph{Warp} and \emph{NSpur}
  probe the periphery of the warped disc on opposite sides of the
  galaxy.
\item The \emph{NE Shelf} samples a major shelf-like over-density in
  the north-east side of the disc.  \cite{fer05} have shown that the
  CMD morphology of the NE Shelf and a region of the GSS bear striking
  resemblance. Indeed, N-body simulations by \citet{fard07} suggest
  that both the Western Shelf (see below) and the NE Shelf might be
  debris from a forward wrap of the GSS progenitor.
\item The \emph{EC1\_field} samples a section of the diffuse Western
  Shelf feature of the Northern minor axis, near the old extended
  cluster EC1 \citep{tan12}. This vast structure was first studied in
  detail by \citet{far12}.
\item The \emph{GC6\_field} and \emph{Brown-disk} sample portions of
  the outer disc of M31.
\item The \emph{Giant Stream} and \emph{Brown-stream} fields probe the
  eponymous GSS of tidal debris falling into the far side of M31. Over
  100~kpc in length \citep{iba07} this accretion event dominates the
  inner halo of M31 and may be related to the North-Eastern and
  Western Shelves.
\item The two fields lying along the Southern minor axis, the
  \emph{Minor Axis} and \emph{Brown-spheroid}, were chosen to probe
  the underlying halo of M31. They were selected because it was
  initially believed that they were free from substructure. However,
  subsequent wide-field mapping as well as the analysis of
  \cite{ric08} showed that both fields are significantly contaminated
  by GSS material.
\end{itemize}

\begin{figure*}
\includegraphics[width=15cm]{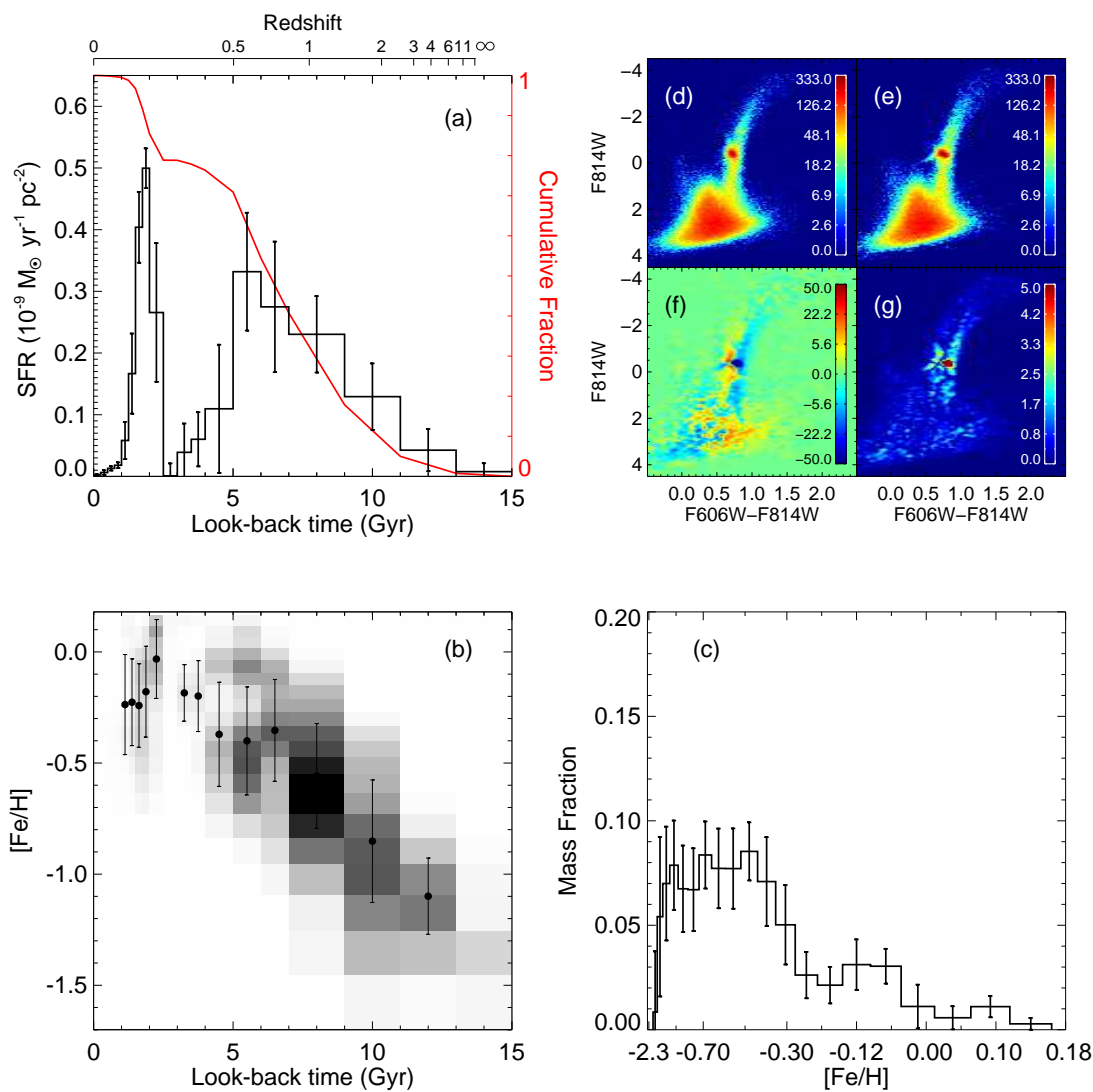}
\caption{Best-fit SFH solution for the {\it Warp}. In the
  counter-clockwise direction, the panels show: (a) the SFR as a function
  of time, normalised to the area of the ACS field (deprojected for the
  M31 disc inclination), (b) the AMR, where the grayscale is proportional
  to the {\it stellar mass} formed in each bin, (c) the metallicity
  distribution of the mass of stars formed, (d)-(e) the Hess diagrams
  (i.e.\ star count per bin) of the observed and best-fit model CMDs,
  (f) the residuals, and (g) the significance of the residuals in
  Poisson $\sigma$. The cumulative mass fraction is shown in red in
  panel (a). The filled circles and error bars in panel (b) show the
  median metallicity and standard deviation in age bins representing
  at least 1~percent of the total mass of stars formed. See text for
  details.}
\label{fig:dl}
\end{figure*}

\subsection{Photometry and artificial star tests}\label{red}

For homogeneity with the analysis presented in \citet{ber12}, we redid
the photometry and artificial star tests following identical methods.%
\footnote{the photometric catalogue, artificial star tests, and
output of the SFH calculations for each field are available on request
to the authors.}
The only difference here is the point spread functions (PSFs) that
were used.  Due to a bug in the {\sc CALACS} processing pipeline, the
pixels affected by cosmic rays were not flagged in the data quality
arrays, which complicated the creation of the model PSF for each ACS
chip image.  Instead, we created PSFs by stacking the 8 (12) images in
F606W (F814W) of the {\it Warp} field data of \citet{ber12} and
selecting over 200 isolated stars per stacked image.  The residuals in
the PSF-subtracted images obtained with these PSFs are significantly
smaller than when using PSFs created from the images affected by
cosmic rays.

The stellar photometry was carried out on the individual exposures with
the standard {\sc daophot/allstar/allframe} suite of programs \citep{ste94}.
The catalogues were then corrected for foreground reddening using the
extinction maps of \cite{sch98}; no further correction
was necessary since the resulting CMDs showed no evidence of
differential reddening. They were also cleaned of non-stellar objects by
applying cuts on the photometric parameters given by {\sc allframe},
namely the photometric uncertainty ($\sigma \leq 0.3$) and the
sharpness, describing how much broader the profile of the object appears
compared to the profile of the PSF ($|{\tt SHARP}|\leq 0.3$).  Finally, we
converted the observed CMDs to absolute magnitudes -- as needed to
calculate the SFH -- using the approximate distance obtained from the
mean magnitude of the red clump (RC) stars. We emphasize the fact that
the SFH calculations do not require a precise distance, as the
algorithm minimizes the impact of the distance uncertainties by shifting the
observed CMD with respect to the synthetic CMD to find the best solution.

\begin{figure*}
\includegraphics[width=15cm]{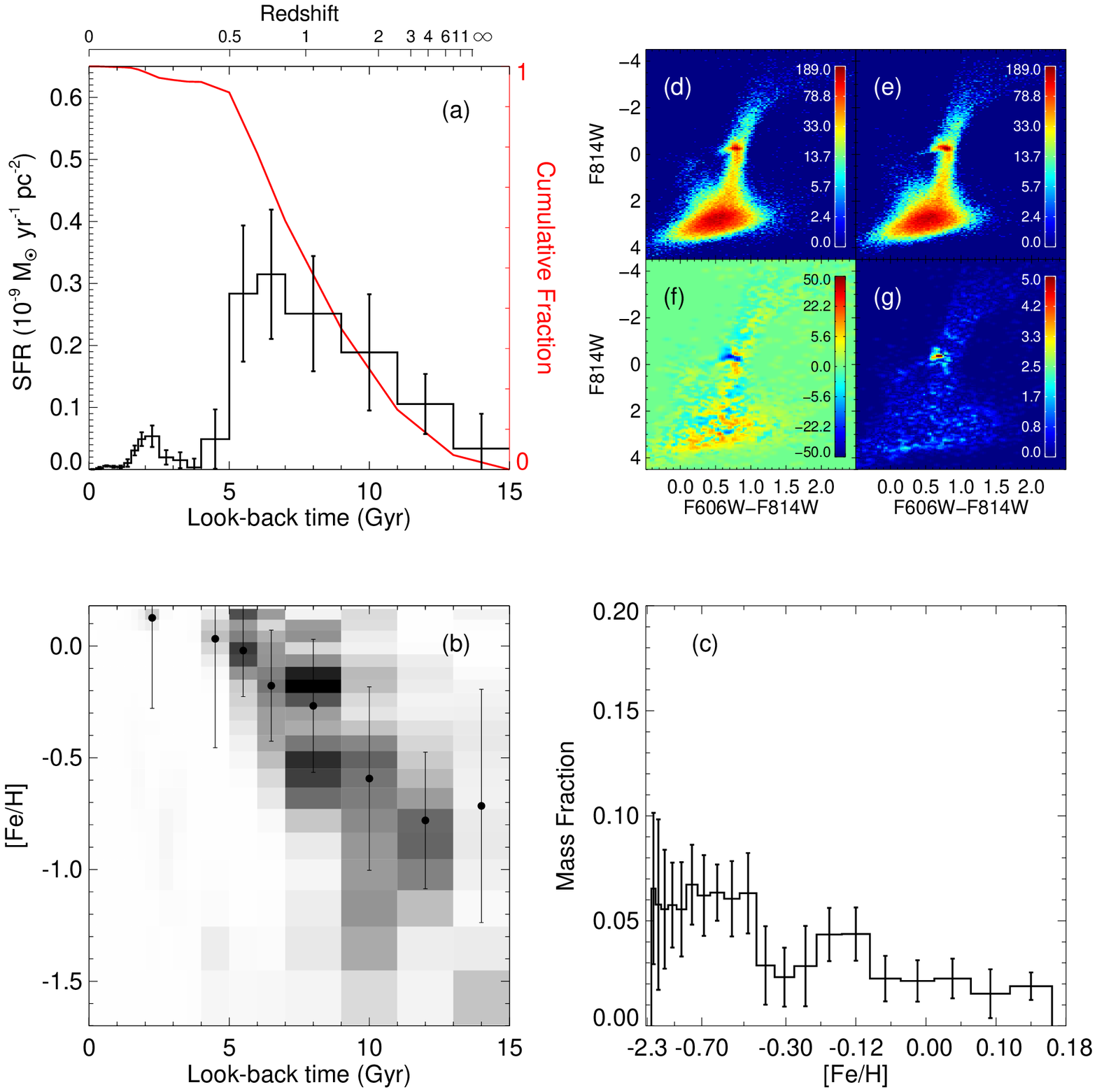}
\caption{Same as Figure~\ref{fig:dl}, but for the {\it Giant Stream}.}
\label{fig:sl}
\end{figure*}

Sample CMDs are shown in the left and middle panels of
Figure~\ref{fig:cmd}, where isochrones and a zero-age
horizontal-branch (ZAHB) from the BaSTI stellar evolution library
\citep{pie04} have been overplotted. The typical exposure times per
pointing of $\sim$2\,400~s in $F606W$ and $\sim$5\,200~s in $F814W$
allowed us to obtain a signal-to-noise ratio $\geq 3$ over three
magnitudes fainter than the horizontal-branch and red clump, i.e.\
comparable to the luminosity of a 12.5~Gyr old MSTO. The CMD of the
{\it Warp}, shown in the middle panel of Figure~\ref{fig:cmd}, is one
magnitude shallower than the CMD based on the full (i.e.\ 10~orbits)
dataset for this field \citep[see Figure~2 of][]{ber12}; this has a
slight effect on the accuracy of our SFHs, in particular at older ages
(see Section~\ref{sec:lit}).

A detailed comparative analysis of the CMDs, including a description
of the main features and motivation for the classification as
`disc-like', `stream-like', or `composite', is given in \cite{ric08}.
Specifically, they note that the stream-like fields harbor a more
prominent blue HB and a wider RGB than the disc-like fields. These
fields also have far fewer main-sequence (MS) stars younger than a
few billion years compared to the disc-like fields. In particular,
they lack the prominent over-density at $M_{606}-M_{F814}=0.5$ and
$M_{814}\sim2$ corresponding to a 2~Gyr old population. The CMDs of
the `composite' fields, on the other hand, have properties which do
not fit simply into either disc-like or stream-like categories.

%%%%%%%%%%%%%%%%%%%%%%%%%%%%%%%% TAB 2 %%%%%%%%%%%%%%%%%%%%%%%%%%%%%%%%%%%%%%%%%
\begin{table*}
 \begin{minipage}{170mm}
  \caption{Results summary for the best-fit SFHs.}
\label{tab:teramo}
% \begin{tabular}{@{\extracolsep{6pt}}cccccccc}
\begin{tabular}{ccc@{\extracolsep{5pt}}c@{\extracolsep{6pt}}c@{}cccc}
\hline 
  Field & Type\footnote{S: stream-like; D: disc-like; C: composite.} & \multicolumn{2}{c}{Stars ever formed} & \multicolumn{2}{c}{Stars still alive} &
  \multicolumn{3}{c}{Percentage of mass formed}\\
  \cline{3-4} \cline{5-6} \cline{7-9}
  name & & $<\!\mbox{age}\!> $\footnote{Median age and metallicity.} (Gyr) & $<\![\rm{Fe/H}]\!>^b $ & $<\!\mbox{age}\!>^b $ (Gyr) & $<\![\mbox{Fe/H}]\!>^b $ & $< 3$~Gyr & $< 5$~Gyr &  $> 8$~Gyr  \\
  \hline 
Giant Stream   & S & 7.9 & $-$0.33 & 7.1 & $-$0.35 & \ 3.4 $\pm$ 0.9 & \ 6.5 $\pm$ 7.7 & 48.3 $\pm$ \ 6.5 \\
Brown-stream   & S & 8.2 & $-$0.35 & 7.1 & $-$0.37 & \ 4.9 $\pm$ 1.8 &  10.2 $\pm$ 8.4 & 51.8 $\pm$ \ 6.1 \\
Minor Axis     & S & 9.7 & $-$0.41 & 8.2 & $-$0.42 & \ 3.9 $\pm$ 4.1 & \ 7.1 $\pm$ 7.2 & 73.7 $\pm$  11.0 \\
EC1\_field     & S & 8.4 & $-$0.52 & 7.0 & $-$0.47 & \ 6.2 $\pm$ 2.0 & \ 8.9 $\pm$ 6.8 & 55.7 $\pm$ \ 7.9 \\
NE Shelf       & S & 7.5 & $-$0.29 & 6.8 & $-$0.40 & \ 3.7 $\pm$ 0.9 &  11.1 $\pm$ 5.7 & 43.0 $\pm$ \ 7.1 \\
 \hline 
{\bf Average}  & S & 8.3 & $-$0.38 & 7.2 & $-$0.40 & \ 4.4 $\pm$ 1.9 & \ 8.8 $\pm$ 7.2 & 54.5 $\pm$ \ 7.7 \\
 \hline 
G1 Clump       & D & 6.6 & $-$0.29 & 5.3 & $-$0.37 &  13.0 $\pm$ 2.6 &  29.3 $\pm$ 4.2 & 30.5 $\pm$ \ 3.7 \\
Warp           & D & 6.3 & $-$0.43 & 4.6 & $-$0.36 &  21.1 $\pm$ 1.8 &  29.1 $\pm$ 3.6 & 29.4 $\pm$ \ 2.1 \\
Claw           & D & 8.3 & $-$0.29 & 6.6 & $-$0.32 & \ 8.0 $\pm$ 2.3 &  19.1 $\pm$ 4.3 & 52.8 $\pm$ \ 4.1 \\
N Spur         & D & 6.8 & $-$0.20 & 5.6 & $-$0.28 &  10.7 $\pm$ 1.8 &  25.7 $\pm$ 3.2 & 34.1 $\pm$ \ 3.6 \\
NE Clump       & D & 7.0 & $-$0.32 & 6.0 & $-$0.40 & \ 7.3 $\pm$ 2.1 &  22.4 $\pm$ 4.6 & 38.0 $\pm$ \ 5.7 \\
 \hline 
{\bf Average}  & D & 7.0 & $-$0.31 & 5.6 & $-$0.35 &  12.0 $\pm$ 2.1 &  25.1 $\pm$ 4.0 & 37.0 $\pm$ \ 3.8 \\
 \hline 
GC6\_field     & C & 8.2 & $-$0.24 & 5.8 & $-$0.26 & \ 8.9 $\pm$ 2.8 &  21.8 $\pm$ 3.1 & 52.0 $\pm$ \ 4.3 \\
NGC205 Loop    & C & 8.5 & $-$0.37 & 6.5 & $-$0.39 & \ 8.8 $\pm$ 3.6 &  18.1 $\pm$ 5.1 & 57.0 $\pm$ \ 5.6 \\
Brown-disk     & C & 7.6 & $-$0.27 & 6.6 & $-$0.32 & \ 5.9 $\pm$ 1.9 &  14.7 $\pm$ 5.1 & 44.8 $\pm$ \ 5.2 \\
Brown-spheroid & C & 8.8 & $-$0.53 & 7.6 & $-$0.50 & \ 4.7 $\pm$ 1.6 & \ 7.6 $\pm$ 6.0 & 59.5 $\pm$ \ 6.2 \\
 \hline 
{\bf Average}  & C & 8.3 & $-$0.35 & 6.6 & $-$0.37 & \ 7.1 $\pm$ 2.5 &  15.5 $\pm$ 4.8 & 53.4 $\pm$ \ 5.4 \\
\hline\vspace{-9mm}
\end{tabular}
\end{minipage}
\end{table*}

%%%%%%%%%%%%%%%%%%%%%%%%%%%%%%%%%%%%%%%%%%%%%%%%%%%%%%%%%%%%%%%%%%%%%%%%%%%%%%%%

%-------------------------------------------------------------------------------

\section{Star Formation Histories}\label{sfh}

\subsection{Method}\label{method}

To build on the results of \citet{ric08}, we have used the technique
of synthetic CMD fitting to provide a more detailed and robust
quantitative analysis of these fields. This involved fitting the
observed data with synthetic CMDs to extract the linear combination of
simple stellar populations (SSP) -- i.e.\ each with small ranges of
age ($\leq$2~Gyr) and metallicity ($<$0.25~dex) -- which provide the
best fit; the amplitudes of which give the rates of star formation as
a function of age and metallicity. When applied to sufficiently
deep CMDs, this technique is very robust and has been shown to produce
virtually indistinguishible results when using different algorithms
\citep[e.g.][]{mon10,hid11}.

We adopted the same methodology as in \citet{ber12} and refer the
interested reader to this paper for a detailed description.  The
method relies on the following programs: {\sc iac-star} \citep{apa04}
to generate the synthetic CMDs against which the observed CMD will be
compared, {\sc iac-pop} \citep{apa09} to find the combination of
synthetic CMDs that best reproduce the observed CMD, and {\sc minniac}
\citep{hid11} to produce the input files for, and process the output
files from, {\sc iac-pop}, and to estimate the uncertainties.

The synthetic CMD from which we extracted the SSP CMDs is based on the
BaSTI stellar evolution library \citep{pie04}; we note that using
a different library only leads to small systematic differences of the
order of $\sim$0.2~dex in metallicity, and $<$1~Gyr in age for ages
older than $\sim$8~Gyr \citep[e.g.][]{mon10,ber12}. The CMD, containing
10$^7$ stars, was generated with a constant star formation rate (SFR)
over wide ranges of age and metallicity: 0 to 15~Gyr old
and $0.0004 \le \rm{Z} \le 0.03$ (i.e. $-$1.7~$\le$~[Fe/H]~$\le$~0.18,
assuming $Z_{\sun}=\rm{0.0198}$; \citealt{gre93}). We adopted a
\citet{kro02} IMF, and assumed a binary fraction of 50~percent and a
mass ratio $q>$~0.5 (see e.g.\ \citealt{kro91,duq91,gal99}).
Given the small size of the ACS field-of-view, the number of
foreground and background contaminants in the regions of the CMD that
are used for the SFH fitting is negligible; our model therefore does
not attempt to account for foreground contamination. Finally,
the incompleteness and photometric errors due to the observational
effects have been simulated for each {\it HST} pointing based on the
results of the corresponding artificial star tests.

The comparison between the observed and synthetic CMDs is performed
using the number of stars in small colour-magnitude boxes. Due to both
observational effects (e.g.\ signal-to-noise ratio) and theoretical
uncertainties in stellar evolution models, some areas of the CMDs are
more reliable than others.  We limit the comparison to a number of
specific areas \citep[called `bundles';][see also
\citealt{mon10}]{apa09} shown as black solid lines in the right panel
of Figure~\ref{fig:cmd}. The bundles are then divided uniformly into
small boxes, with a size depending on the density of stars and
reliability of the stellar evolution models: these are shown by thin
blue lines in the right panel of Figure~\ref{fig:cmd}.
Since every box carries
the same weight, the number of boxes in a bundle determines the weight
that the region has for the derived SFH. In our analysis, the main
bundle covers the MS and sub-giant branch, where the models are less
affected by the uncertainties in the input physics; this bundle is
divided into 0.1$\times$0.06~mag boxes.  The other three bundles we
use lie at the base, on the red side, and over the tip of the RGB and
serve to provide mild constraints on the metallicity.  Since the
models for these evolved phases are more uncertain, and the
differences between stellar libraries more important \citep{gal05}, we
used significantly larger boxes; 0.1$\times$0.5 for the former, and
0.3$\times$0.5 for the latter two. This yields a total number of
$\sim530$ boxes.

The best-fit SFH is determined by finding the amplitudes of the linear
combination of SSP CMDs which best match the observed CMD. The number
of observed stars, and artificial stars from each SSP, counted in each
CMD box, serves as the only input to {\sc iac-pop}.  No a priori AMR,
or constraint thereon, is adopted: {\sc iac-pop} solves for both ages
and metallicities simultaneously within the age-metallicity space covered
by our SSPs. The goodness of the coefficients in the linear combination
is measured through the modified $\chi^2$ statistic of \citet{mig99},
which {\sc iac-pop} minimizes using a genetic algorithm. These
coefficients are directly proportional to the star formation rate of
their corresponding SSPs.

To sample the vast parameter space, the $\chi^2$ minimization is
repeated several hundred times for each field after shifting the bin
sampling in both colour--magnitude and age--metallicity space. In
addition, the observed CMD is shifted with respect to the synthetic
CMDs in order to account for uncertainties in photometric zero-points,
distance, and mean reddening.

Finally, the uncertainties on the SFRs
were estimated following the prescriptions of \citet[see also
\citealt{apa09}]{hid11}. The
total uncertainties are assumed to be a combination (in quadrature) of
the uncertainties due to the effect of binning in the colour--magnitude
and age--metallicity planes, and those due to the effect
of statistical sampling in the observed CMD.

\subsection{Results}\label{results}

Our results reveal quantitative differences in the derived SFHs that
fully support the inferences made by \citet{ric08} on the basis of a
comparative analysis of CMD morphology alone.  Rather than showing the
detailed best-fit solutions for each of the fourteen fields, we
present solutions for an illustrative disc-like and a stream-like
field, as well as the main features in synthesized form for all the
fields; the individual solutions for the other fields are given in the
Appendix. We also summarize the results in Table~\ref{tab:teramo}: for
each field, we list the median age and metallicity of all the stars
ever formed, as well as the percentage of the total stellar mass
formed in the last 3~Gyr and 5~Gyr ($z\sim0.5$), and the mass already
in place by $z\sim1$ (8~Gyr ago) according to the best-fit solutions.
To allow comparison with observations, we include the median age and
metallicity of the stars that are still alive today (i.e.\ excluding
the stars that have evolved to become white dwarfs or supernovae) in
columns 5 and 6.  We also show the average quantities for each field
type.

\begin{figure}
\includegraphics[width=8.3cm]{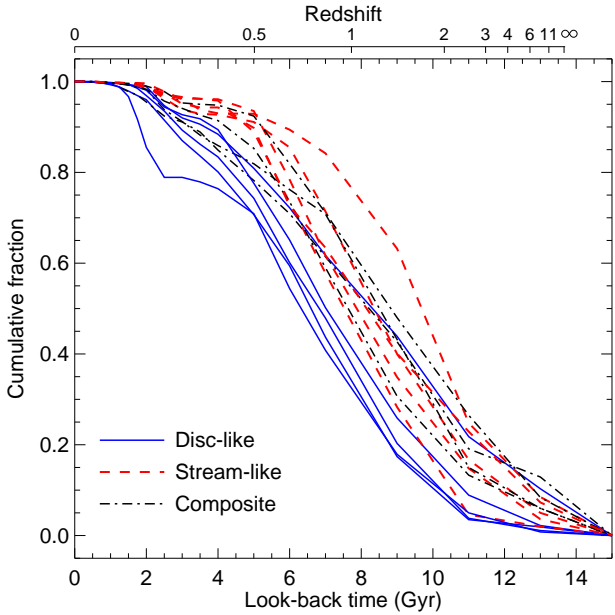}
\caption{Cumulative mass distributions for our 14 fields showing the fraction
  of stellar mass formed in each of the substructure fields as a function of
  look-back time. The disc-like, stream-like, and composite fields are
  represented by solid blue lines, red dashed lines, and black dotted lines,
  respectively. The disc-like fields can be seen to be systematically younger,
  on average, than the other fields.}
\label{fig:cf}
\end{figure}

Figures~\ref{fig:dl} and \ref{fig:sl} show the best-fit solutions for
a disc-like field (the {\it Warp}) and a stream-like field (the {\it
Giant Stream}), respectively.  In the counter-clockwise direction,
the panels show: the evolution of the SFR (a) and metallicity (b) as a
function of look-back time; (c) the metallicity distribution of the
mass of stars formed; (d)-(e) the Hess diagrams of the observed and
best-fit model CMDs, with a logarithmic stretch to bring out fainter
features; (f) the residual differences between the two, in the sense
observed$-$model; and (g) the absolute residual difference normalized
by Poisson statistics in each bin. The redshift scale shown on panel
(a) was constructed assuming the WMAP7 cosmological parameters from
\citet{jar11}, namely $H_0=71.0\ \rm{km}\ \rm{s}^{-1}\ \rm{Mpc}^{-1}$,
$\Omega_\Lambda=0.73$ and $\Omega_{\rm{M}}=0.27$.

\begin{figure}
\includegraphics[width=8cm]{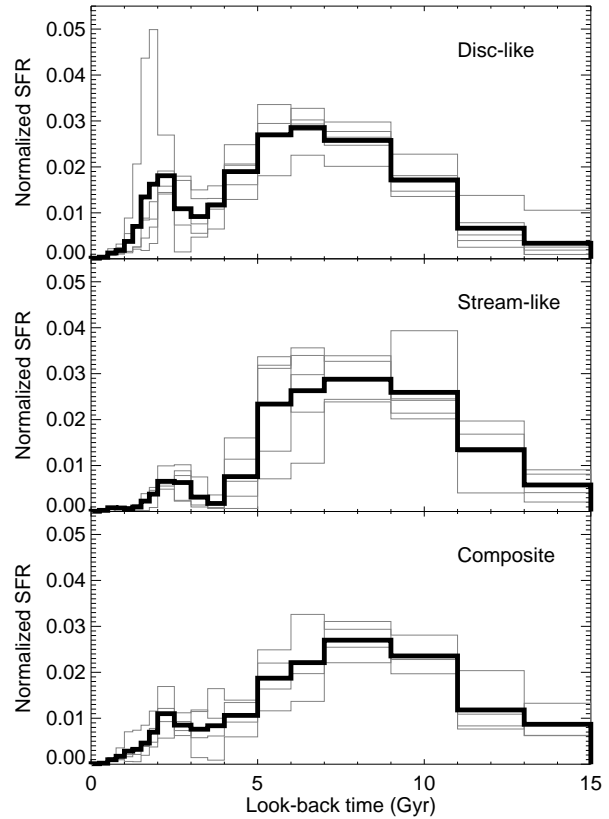}
\caption{The SFH of the disc-like (top), stream-like (middle), and composite
 (bottom) fields, normalised to the total mass of stars formed in each field,
 where the individual fields are shown with different line styles. The thick
 gray lines represent the average of the normalized SFHs.}
\label{fig:sfhs}
\end{figure}

The comparison of panels (d) and (e) in these figures shows that
the best-fit SFH solutions reproduce the data very well. The panels
showing the residuals of the fits indicate where the models struggle
to replicate the data -- significant deviations of the model from the
data appear as strong coherent residuals.  The RC is the only phase
of stellar evolution that the algorithm has consistently failed to
fit well due to known limitations of the current stellar evolution
models \citep{gal05}. As discussed previously, this is the reason why
this region of the CMD has not been included in the SFH calculations,
and therefore carries no weight in our final solution.  Otherwise the
patterns of residuals display no significant systematic disparities
between the models and the data;
while the model RGB close to the tip is sometimes wider than the
observed RGB, we recall here that the theoretical models of this
evolutionary phase are less reliable \citep[e.g.][]{gal05,cas13}, hence
their low weight in the SFH calculations. Besides, none of the fields
have residuals in this part of the CMD with significance larger than
$\sim$1-$\sigma$.

Our homogeneous analysis of the best-fit SFH results demonstrate a
dichotomy in the nature of the stellar substructure in M31's inner
halo, as first pointed out by \citet{ric08}.  The main differences
between the two types of fields are the typically older median ages of
the stream-like fields, as well as their more rapid chemical
enrichment. The former is illustrated in Figure~\ref{fig:cf}, where
the cumulative mass fractions of all fourteen fields are shown
together. It shows that the disc-like fields, represented by solid
blue lines, formed very few stars before about 10~Gyr ago ($z \sim
2$), and have a median age of $\sim$7~Gyr. Instead, star formation
started very early in the stream-like fields but decreased to residual
levels by 4--5~Gyr ago; their median age is on average 1.5~Gyr older.
As expected from their CMD morphologies, the fields flagged as
composite in \citet{ric08} have properties intermediate between the
other two: an early onset of star formation like the stream-like
fields, yet substantial star formation in the past 5~Gyr (see below).

\begin{figure}
 \includegraphics[width=8.3cm]{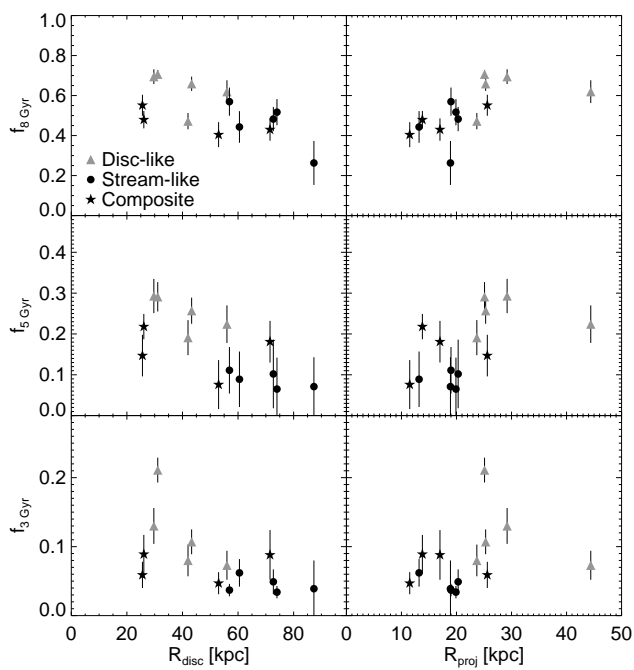}
 \caption{Fraction of stars formed in the last 8~Gyr (top), 5~Gyr (middle),
 and 3~Gyr (bottom) as a function of de-projected (left) and projected
 (right) radius. The disc-like, stream-like, and composite fields are
 represented by gray triangles, black filled circles, and black stars,
 respectively. Note the differing vertical scales.}
\label{fig:sfh_prof}
\end{figure}

Figure~\ref{fig:sfhs} shows the SFR as a function of time for all the
fields, separated by field type and normalized to the total mass of
stars formed in each field, as well as the mean SFR for each type.  In
\citet{ber12}, we found that the {\it Warp} underwent a brief but
strong burst of star formation $\sim$2~Gyr ago, which we suggested
could have been triggered by the pericentric passage of M33 at that
time. Interestingly, Figure~\ref{fig:sfhs} shows that {\it all} our
fields show an enhancement in the star formation rate at this time,
although the relative strength of this burst seems stronger in the
disc-like fields. We note that this young population is unlikely to be
due to blue stragglers, since it represents a significant fraction of
the total mass of stars formed (between 5 and 25~percent).  Moreover,
these stars have a higher metallicity than expected from merging or
mass-transfer in old, metal-poor stars, the currently-favoured
scenarios for producing blue stragglers.  We discuss the implications
of detecting this apparently ubiquitous burst in
Section~\ref{sec:discdl}.

\begin{figure}
\includegraphics[width=7.5cm]{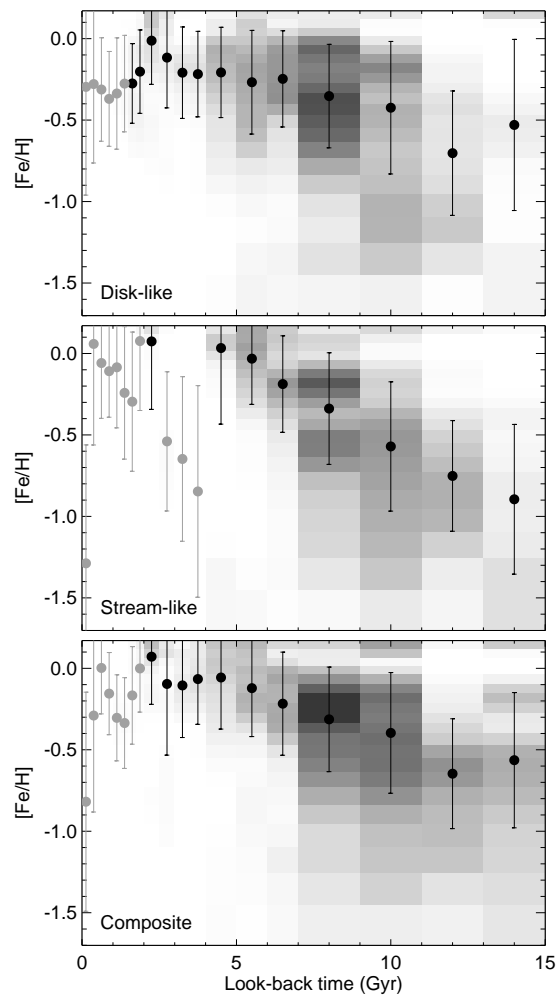}
\caption{The AMR of the disc-like (top), stream-like (middle), and
  composite (bottom) fields, where the grayscale is proportional to
  the stellar mass formed in each bin. The AMRs of the individual
  fields were weighted by the total mass of stars formed in each.
  Filled circles mark the median metallicity in each age bin; gray
  symbols were used for age bins representing less than 1~percent of
  the total mass, where the metallicity is therefore less reliable.
  Error bars show the standard deviation in each bin.}
\label{fig:cels}
\end{figure}

Figure~\ref{fig:sfh_prof} shows the fraction of the total stellar mass
formed in the last 8~Gyr ($z \sim 1$, top), 5~Gyr ($z \sim 0.5$,
middle), and 3~Gyr (bottom) as a function of the deprojected (left)
and projected (right) radius from the centre of M31.  The disc-like,
stream-like, and composite fields are represented by gray triangles,
dark gray circles, and black stars, respectively. The trend in the top
left panel indicates that the outermost fields had a significant
fraction of the mass already in place by $z \sim 1$, while the fields
closer to the centre of M31 are on average younger. However, this may
just reflect the complicated merger history of this galaxy rather than
imply an outside-in formation of the outer disc and inner halo. In
addition, the deprojected distance is based on the assumption that all
the fields are located in the same plane as the M31 disc, which is
clearly not always the case \citep[e.g.][]{mcc03}. The top right panel
shows the reverse trend -- i.e. younger populations at larger radii --
but this simply reflects the locations of the fields probed: the
disc-like (stream-like) fields tend to be located close to the major
(minor) axis, thus appearing farther from (closer to) the centre of
M31.

\begin{figure}
\includegraphics[width=8cm]{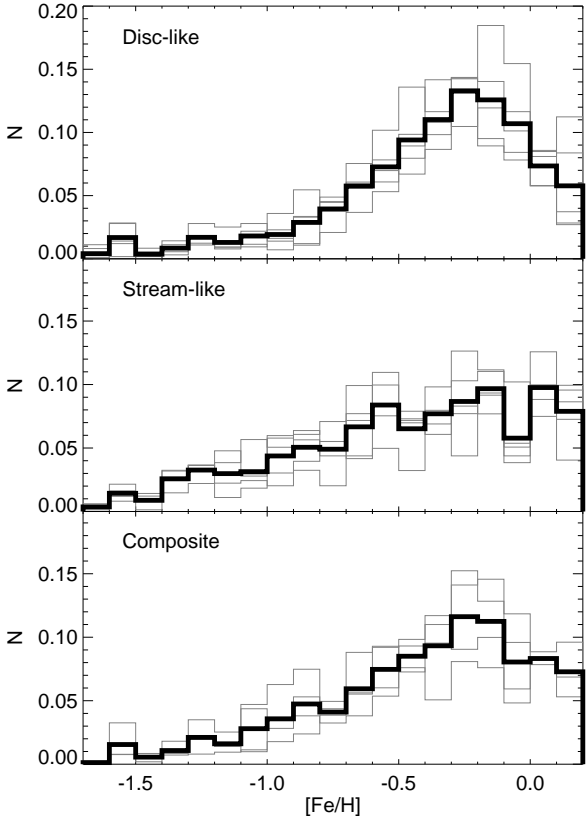}
\caption{The MDFs of stars brighter than $M_{814}<-1.5$ that are still
  alive today, according to our best-fit SFHs, in the disc-like (top),
  stream-like (middle), and composite (bottom) fields. These are
  normalized to the total mass of stars formed in each field. The line
  styles are as in Figure~\ref{fig:sfhs}.}
\label{fig:mdfs}
\end{figure}

Indeed the middle panels of Figure~\ref{fig:sfh_prof} indicate that the
disc-like fields have a systematically larger fraction of young stars
than the stream-like fields: disc-like fields formed $\sim$25~percent
of their stellar mass in the last 5~Gyr while stream-like fields formed
only about 9~percent in the same time period.  Similarly, in the past
3~Gyr, the disc-like fields formed $\sim$12~percent of their total
stellar mass, compared to only 4~percent for the stream-like fields.
Overall, Figure~\ref{fig:sfh_prof} shows that there are no strong
radial gradients in the median age of the stellar population at these
galactocentric distances.

The other significant difference between the disc-like and stream-like
substructure fields is evident in their AMRs. Figures~\ref{fig:dl} and
\ref{fig:sl} show that while the metallicity evolution is rather well
constrained in both cases, it proceeded at a different rate. The
metallicity of the {\it Giant Stream} covers the whole range spanned
by the isochrone set, and reached the maximum metallicity of the grid
roughly 5~Gyr ago. On the other hand, the metallicity evolution was
slower in the {\it Warp}, lacking a population of the most metal-poor
stars and reaching solar metallicity only $\sim$2~Gyr ago.

The AMRs of the other stream-like fields are basically
indistinguishable from that of the {\it Giant Stream} field,
suggesting a high degree of chemical homogeneity in the stream
progenitor. While the AMRs of the disc-like fields show somewhat more
variation, they share the same features as the AMR of the {\it Warp}
-- i.e.\ generally lacking the two extremes of the metallicity range,
leading to a milder metal enrichment.  Some of these fields are
located on rather diffuse and complex structures in the outskirts of
M31, so the complexity of their AMRs may be a consequence of the
superposition of substructure along the line of sight.

Figure~\ref{fig:cels} presents the weighted mean AMRs of the disc-like,
stream-like, and composite fields, where the median metallicity and
standard deviation in each age bin are shown as filled circles and
error bars. The difference in chemical enrichment histories persists
here. The top panel shows that the metallicity of the disc-like fields
increased from [Fe/H]$\sim-0.7$ to solar metallicity $\sim$2~Gyr ago.
The decline in global metallicity following the recent burst was
already observed in our SFH of the {\it Warp} obtained from the full
depth data, where we have shown that it is not an artifact of the
method \citep[][see also \citealt{bro06}]{ber12}. In contrast, the AMR
of the stream-like fields reveals a more rapid chemical enrichment,
from [Fe/H]$\sim-1$ to solar metallicity before 5~Gyr ago.

The difference in AMRs is also reflected in the predicted present-day
metallicity distribution functions (MDFs). Figure~\ref{fig:mdfs} shows
the MDFs of stars brighter than M$_{814}<-1.5$ (i.e.
F814W~$\sim23$\footnote{This is roughly the magnitude limit at which
spectroscopy of individual stars can be carried out with the current
generation of 8-m telescopes and instruments.})  {\it that are still
alive today} according to the best-fit SFHs for each field. The
MDFs were obtained by selecting the bright stars from the solution
CMDs (panel (e) in, e.g.\ Figure~\ref{fig:dl}), for which we know
the individual age and metallicity. We find
that the MDF of the disc-like fields has a bell-shape peaking at
[Fe/H]$\sim-$0.2, with very few low-metallicity stars. On the other
hand, the stream-like MDF is significantly flatter, and although
increases smoothly it has no clear peak. The fact that considerable
stars are present in the highest metallicity bin available in our
models may indicate that even more metal-rich stars are present in the
GSS.

Finally, the `composite' fields display a variety of behaviours which
do not fit easily into either disc-like or stream-like categories,
consistent with them being a complicated mixture of both. The SFH of
the {\it Brown-spheroid} field is more akin to those of the
stream-like fields, while the three other composite fields more
resemble the disc-like fields.  Their SFRs (Figure~\ref{fig:sfhs}) and
MDFs (Figure~\ref{fig:mdfs}) are intermediate between those of the
stream-like and disc-like fields, while their AMRs are closer to the
latter fields.  It is likely that these fields sample regions where
material stripped from the GSS progenitor is mixed together with
material from the perturbed thin disc, or that the GSS progenitor
contained gradients in its stellar populations. For example, if the
progenitor was a disc galaxy \citep[e.g.][]{far08} it would have shed
different populations (corresponding to mixtures of bulge, disc and
halo stars) as it dissolved in its orbit around M31.  We do not
discuss these fields further and instead focus on the prominent
differences between disc-like and stream-like fields and what they
imply for the origin and nature of the underlying populations.

%-------------------------------------------------------------------------

\section{Discussion and Interpretation}\label{interp}

\subsection{Comparison with literature results}\label{sec:lit}

The average metallicities of all the fields, shown in
Table~\ref{tab:teramo}, lie in the range $-0.5 \la [\mbox{Fe/H}] \la
-0.3$. This is in good agreement with the previous photometric and
spectroscopic studies probing the outer disc and inner halo of M31, as
well as the GSS \citep[e.g.][]{mou86,dur94,hol96,fer01,bel03,fer05,
bro06,kal06,gil09,far12,iba14}.

There are a few papers in the literature in which the SFH of a subset
of the fields presented here was calculated using similar methods.
Using {\sc StarFISH} \citep{har01}, \cite{bro06} fit the SFHs of the
{\it Brown-stream}, {\it Brown-spheroid}, and {\it Brown-disk}, at
their full depth, $\sim$1.5~mag deeper than in this analysis and
consisting of over 32 {\it HST} orbits per pointing.  We have analysed
only a subset of these data in order to match the depth of our other
HST/ACS fields and thus allow a homogeneous comparison. The additional
depth used in the Brown et al. analysis had the advantage of allowing
them to reach the oldest MSTO, enabling a probe of the earliest stages
of star formation with more detail and accuracy than we have here.
Although following a similar synthetic CMD technique to the one
described here, \cite{bro06} used the Victoria-Regina isochrones
\citep{van06} with a larger range of metallicities ($-2.2 \leq
\mbox{[Fe/H]} \leq +0.5$), so minor systematic differences are
expected.

\begin{figure*}
\includegraphics[width=8cm]{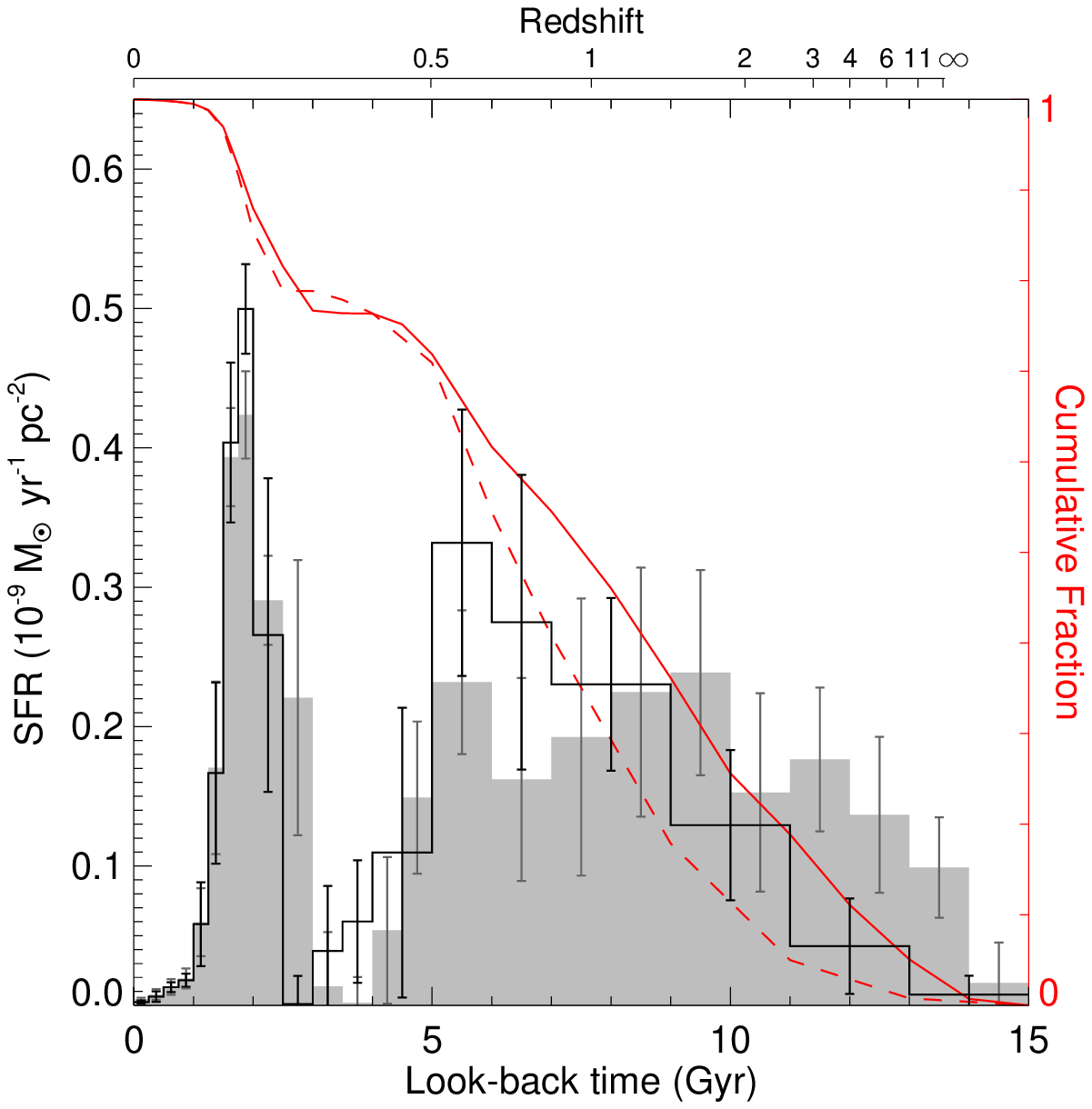}
\includegraphics[width=8cm]{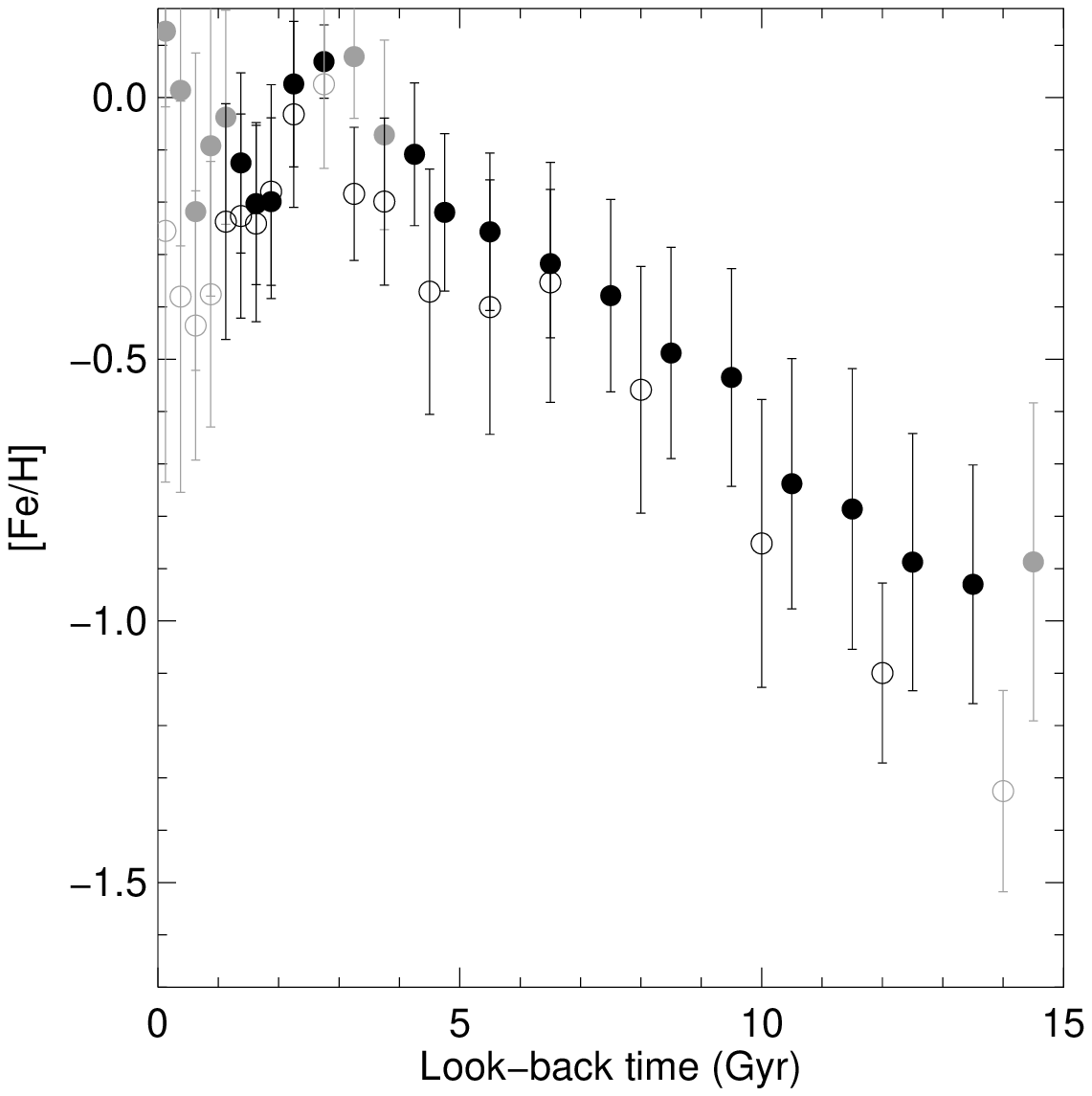}
\caption{Comparison of the {\it Warp} SFH obtained with the deep dataset
 in \citet{ber12} and with the shallower data in this work (see text for
 details).
 {\it Left}: SFR as a function of time for the deep (filled gray histogram)
 and shallower (black histogram) datasets. The solid (dashed) red line shows
 the cumulative mass fraction of the deep (shallower) dataset.
 {\it Right}: Median metallicity as a function of time. Filled and open
 circles represent the deep and shallower data, respectively. As in
 Figure~\ref{fig:cels}, gray symbols denote age bins that are less reliable
 due to the low SFR. Error bars show the standard deviation in each bin.}
\label{fig:comp}
\end{figure*}

It is difficult to quantitatively compare their results to ours
because they only present qualitative AMRs (similar to our panel (b)
in Figures~\ref{fig:dl} and \ref{fig:sl}), without the corresponding
histograms showing the evolution of the SFR with time, the metallicity
distribution, and their associated uncertainties.  However, the
qualitative agreement between their AMRs (their Figures 9, 13 and 18)
and ours is excellent, especially given the different libraries used
and the range of metallicities available in the isochrone sets.  They
found that the majority of stellar mass in the {\it Brown-stream} and
{\it Brown-spheroid} fields formed between 5 -- 14~Gyr with a
considerable range of metallicities and rapid chemical enrichment. All
three of the fields they analysed showed low amplitude star formation
within the last 4~Gyr, while their {\it Brown-disk} field was found to
be dominated by younger (4 to 8~Gyr old) populations and lacked
metal-poor stars. These findings mirror the results we have obtained
here from an analysis of shallower versions of their datasets.

Secondly, \cite{fari07} also used {\sc StarFISH} to model the MS, RC
and RGB of the {\it G1 Clump} and visually compared the simulated
luminosity functions and distribution of stars on the CMD to the
observed data.  They concluded that the bulk of the mass in the {\it
  G1 Clump} was in place more than 6~Gyr ago but that 10~percent of
the mass was formed in the last 2~Gyr. Based on
Table~\ref{tab:teramo}, we find that $\sim$71~percent of the mass was
in place by 5~Gyr ago and $\sim$13~percent formed in the last 3~Gyr.
\citet{fari07} also find a relatively high mean metallicity of [Fe/H]
= $-0.4$, a large metallicity dispersion of $\ga 0.5$~dex and little
evidence for chemical evolution. These results are in very good
agreement with those presented here.

Finally, in \citet{ber12} we presented the SFH of the {\it Warp} based
on the full dataset (i.e.\ 10~orbits vs.\ the 3 used here), reaching
about one magnitude deeper. The higher signal-to-noise ratio at the
magnitude of the MSTO and larger number of stars in the CMD allowed us
to use a finer grid in age and metallicity \citep[see Figure~6 in][]{ber12}.
The comparison of the results is presented in Figure~\ref{fig:comp}. It
shows that the SFR and AMR obtained in each case are in very good agreement,
with most discrepancies smaller than 1-$\sigma$. We note that the SFH
obtained from the shallower data appears to be slightly offset toward
younger ages ($\sim$0.5 Gyr on average) and lower metallicities (mean
$\Delta$[Fe/H] = 0.12); however, these offsets are smaller than the ones
we found when comparing the SFHs obtained with the BaSTI and Padova
\citep{gir00} libraries on the deep dataset \citep[Appendix~A in][see
also \citealt{bar11}]{ber12}, and therefore are not the dominant source
of uncertainties. The main drawback of the shallower data is that the
details of the SFH at older ages are less reliable \citep[see also][]{wei14}.
In particular, the AMR is not as well constrained, leading to higher
uncertainties on the age of stars beyond $z \sim 1$, and the median age is
found to be slightly younger ($\sim$6.5~Gyr) than from the deeper data
($\sim$7.5~Gyr). According to the 50~percent completeness limits, the
{\it Warp} is one of the shallower fields analyzed here -- the fourth out
of 14. We therefore expect that the limitations due to the depth of the
photometry will be similar or less significant in most of the fields.

\subsection{Constraints on the SFH of the Giant Stream Progenitor}\label{sec:discsl}

The GSS was discovered in the INT/WFC wide-field panoramic survey of
M31 over a decade ago \citep{iba01}, and is falling in from the far
side of M31 \citep{mcc03}. It has been modeled with some success by
various groups \citep[e.g.][]{iba04,fon06,fard07,mor08,sad14}.  These
N-body simulations suggest that the stream and associated shelves are
the tidal debris of a progenitor with a total mass of a few
$10^9-10^{10}M_{\sun}$ that fell in to M31 in the last billion years.
It is inferred to have come in on a highly radial orbit, and have
wrapped around the inner galaxy at least twice. However, the exact
nature of the satellite involved in this substantial accretion event
remains a mystery. Kinematical constraints rule out both M\,32 and
NGC\,205 as candidates and simulations predict that the remnant should
be located in the northeast half of the M31 disc \citep{iba04,far13}.
To date, no obvious source has been found in this region with either
the INT/WFC survey, nor in the deeper {\it Pan-Andromeda
Archaeological Survey} \citep{mcc09}.

As the GSS progenitor fell towards M31, long tidal tails are expected
to be produced, both leading and trailing the progenitor core.
Through comparison to the N-body simulation of \citet{fard07},
\citet{ric08} demonstrated that the {\it Giant Stream} and {\it
Brown-stream} fields directly probe the trailing stream while the
{\it NE Shelf}, {\it EC1\_field} and {\it Minor Axis} fields probe
wraps of the leading stream.  The results of the SFH fits allow us to
place new constraints on the nature of the GSS progenitor.  Perhaps
the most interesting feature of the best-fit SFHs of all the
stream-like fields is the ubiquitous spread of metallicity ($\ga
1.5$~dex) among the older populations, as expected from the
significantly redder RGB than in the smooth halo component
\citep{fer02, iba14}.  We find that these fields enriched from
$\sim-1.5$~dex to at least solar metallicity within $\sim$8~Gyr of
evolution.  Such rapid enrichment is characteristic of early-type
dwarf galaxies and the bulges of spirals \citep[e.g.\ Sagittarius
dwarf:][]{sie07}.  Interestingly, some properties of the observed
stream -- in particular the asymmetric distribution of stars along the
stream cross section -- are better reproduced in N-body models in
which the progenitor possessed a rotating disc
\citep[e.g.][]{far08,sad14}.  The homogeneous AMRs we have derived for
the stream-like fields suggest that either there were no strong
population gradients in the progenitor or that the inner halo is
littered with material from only the central regions of the
progenitor, rather than the metal-poor stars which might dominate its
outskirts.

Another common feature of all stream-like fields is the sharp decrease
in star formation $\sim$5~Gyr ago, also seen by \cite{bro06} in the
field they analysed.  Additionally, \citet{fer02} noted the lack of
intermediate-age AGB stars in the substructure associated with the
stream.  This quenching of star formation is interesting, as it may
indicate when the progenitor first entered the halo of M31. Ram
pressure by hot gas in the corona of large galaxies is well known to
strip smaller galaxies of their gas and lead to the cessation of star
formation \citep[e.g.][]{may06}. Indeed, a detailed comparison of the
stellar and gaseous distributions around M31 found no \hi\ associated
with the GSS, indicating that whatever gas the progenitor had was lost
a long time ago \citep{lew13}. The results from our SFH fits may help
to tailor future the N-body simulations of the GSS progenitor's
orbital evolution around M31.

\subsection{Ubiquity of the 2-Gyr-old burst: evidence for disc heating?}\label{sec:discdl}

In \citet{ber12}, we demonstrated that the {\it Warp} field underwent
a strong burst of star formation $\sim$2~Gyr ago that lasted about
1.5~Gyr.  Prior to this, there was a rapid decline in the star
formation rate and almost a complete lull in activity for the
preceding Gyr.  In that analysis, we interpreted this burst as a
consequence of the close passage of M33, which self-consistent N-body
modeling indicated should have occured around the time of the onset of
the burst \citep{mcc09}.  Such an interaction and subsequent burst are
further supported by the overabundance of 2~Gyr old star clusters
clusters \citep{fan10} and planetary nebulae \citep{bal13}. However,
our interpretation relied on the SFH of a single pencil-beam pointing
in the outer disc of M31.

The SFH derivations presented here enable us to strengthen the
argument for a large-scale burst of star formation in M31 which took
place $\sim2$~Gyr ago. All 14 fields analysed in this work show
an enhancement of the star formation rate at this epoch, regardless of
their location.  Considerable field-to-field variations exist in the
intensity of this enhancement, however.  After the {\it Warp}, the
{\it NSpur} and {\it GC6\_field} fields exhibit the next most intense
bursts of star formation, both of which reach peak intensities that
are roughly half of that seen in the {\it Warp}.  It is notable that
the {\it NSpur} is located at the opposite end of the M31 disc from
the {\it Warp} (i.e.\ about 50~kpc away).  The detection of the burst
over such a large area, as well as in the outer disc of M33
\citep[see][and references therein]{ber12}, lends further credence to
the idea of a disc-wide burst triggered by a close passage.

The large intensity of star formation during the burst in the {\it
Warp} and {\it NSpur} fields is most likely due to their location
within the thin disc of M31. However, the detection of the 2~Gyr old
burst in the other fields, in particular those at locations associated
with large deprojected radii, is more puzzling. Deep 21~cm
observations show that high column density neutral hydrogen in M31 is
largely confined within the inner 2$\degr$ \citep[27~kpc, shown as the
inner ellipse in Figure~\ref{fig:map};][]{bra09,che09}, while the
molecular hydrogen -- as traced by the carbon monoxide lines -- is
even less extended \citep[${\rm R}_{disc}\la$~16~kpc;][]{nie06}.
Unless the gas distribution in M31 was significantly more extended in
the past, it seems unlikely that star formation could have occured
{\it in situ} in our outermost fields.  Instead, it seems more
plausible that the young populations in these fields originate in
material that has been recently kicked-out from the thin disc.  This
idea was first raised by \citet{ric08} who noticed the similarity
between some inner halo CMDs and the stellar populations in the thin
disc of M31.  There has been mounting evidence over the past two
decades of the disruptive effects of minor mergers and continued
interactions with dark matter sub-halos to heat, thicken and perturb
MW-like stellar discs, providing a method capable of producing such
low-latitude substructure while preserving the global thin disc
structure \citep[e.g.][]{qui93,wal96,vel99,gau06,kaz09,pur10,tis13}.
The fact that some of the M31 substructure exhibits disc-like
kinematics, even at radii of $\sim$70~kpc, certainly supports the idea
that the disc has undergone significant disruption \citep{iba05}.
Additionally, kinematic evidence for the presence of heated disc stars
in the halo has recently been found in both the Milky Way
\citep{she12} and M31 \citep{dor13}.

Hence, we favour an interpretation where the young populations were
formed at smaller galactocentric distances, where the molecular
gas density was sufficient to support star formation, and were
subsequently displaced to their present locations through the
rearrangement of material following a minor merger.  There is ample
evidence that M31 has had a rich interaction and merger history
\citep[e.g.][]{iba01,mcc09,gil09,bat13}, with the most significant
recent event being the accretion of the GSS progenitor.  Recent
modelling work suggests the last pericentric passage of the progenitor
took place $760\pm50$~Myr ago \citep{far13}. This event may
have had enough clout to heat and perturb the thin disc to
redistribute disc stars of all ages to large distances, and is thus
consistent with the fact that a population of $\sim2$~Gyr stars is
present in all fields, regardless of their radial distance or
position relative to the main gas disc.

%-------------------------------------------------------------------------------

\section{Conclusions}\label{cl}

We have presented the quantitative SFHs of 14 deep {\it HST/ACS}
fields which probe different stellar substructures in the outer disc
and inner halo of M31. The SFHs we derive reinforce both the results
obtained from a morphological comparison of their CMDs by
\citet{ric08}, and the results from the analysis of a much deeper CMD
of the {\it Warp} field by \citet{ber12}. In particular, we find that:

\begin{itemize}
\item the classification of the fields into two primary categories --
  `stream-like' and `disc-like' -- still holds when examining their
  quantitative SFHs. We find that disc-like fields formed most of
  their mass ($\sim$65~percent) since $z\sim1$, yielding a median
  age of 7~Gyr, with one quarter of the stellar mass formed since 5~Gyr
  ago.  The stream-like fields, on the other hand, are on average
  1.5~Gyr older and have formed $\la$10~percent of their stellar mass
  within the last 5~Gyr.

\item the stream-like fields are characterised by an AMR showing rapid
  chemical enrichment from [Fe/H]$\sim-$1.7 to roughly solar
  metallicity by $z=1$, suggesting an early-type progenitor such as a
  dwarf elliptical galaxy or a bulge.

\item the burst of star formation about 2~Gyr ago, first identified in
  our previous analysis of the southern warp as well as in the outer
  disc of M33, is detected in {\it all} the fields studied here,
  albeit with varying intensity.  The widespread nature of this event,
  combined with the fact it coincides with predictions for the last
  pericentric passage of M33, lends strong support to the interaction
  hypothesis.

\item the presence of $\sim$2~Gyr old stars over 50~kpc from the
  centre of M31 -- or 10~kpc above the disc, i.e.\ well beyond the
  current extent of both the molecular and neutral hydrogen discs --
  is most easily explained if this population has been scattered from
  the thin disc during a recent encounter.  The close interaction
  between M31 and M33, combined with the recent impact of the GSS
  progenitor, add weight to the idea that repeated interactions
  between M31 and its massive satellites could have enough clout to
  heat and perturb the thin disc to redistribute some disc material
  into the complex substructure we see today.
\end{itemize}

%-------------------------------------------------------------------------------

\section*{Acknowledgments}

This work was supported by a Marie Curie Excellence Grant from the
European Commission under contract MCEXT-CT-2005-025869 and a
consolidated grant from STFC. SLH and AA acknowledge support by the
IAC (grant 310394) and the Science and Innovation Ministry of Spain
(grants AYA2007-3E3507 and AYA2010-16717). GFL thanks the Australian
Research Council for support through his Future Fellowship
(FT100100268) and Discovery Project (DP110100678).

%-------------------------------------------------------------------------------

%-------------------------------------------------------------------------------

\appendix

\section{SFHs of individual fields}

The results of the best-fitting SFHs for the remaining twelve fields are
available as Supporting Information with the online version of the paper:
disc-like fields in Figures~A1--A4; stream-like fields in Figures~A5--A8;
and composite fields in Figures~A9--A12.

\begin{figure*}
\includegraphics[width=15cm]{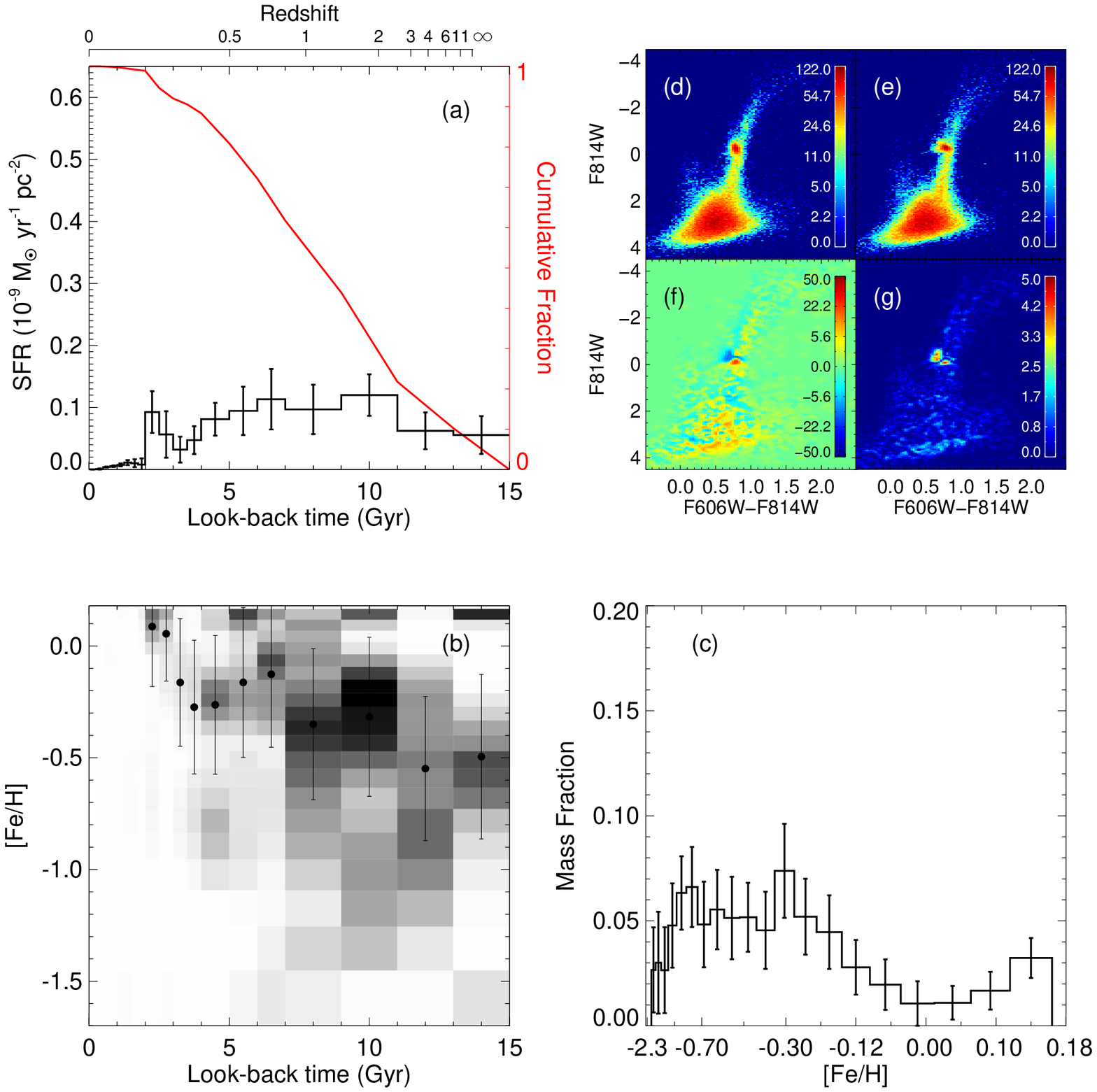}
\caption{Best-fit SFH solution for the {\it Claw}.
 The panels show: (a) the SFR as a function of time, normalised to the
 deprojected area of the ACS field, (b) the AMR, where the grayscale is
 proportional to the {\it stellar mass} formed in each bin, (c) the
 metallicity distribution of the mass of stars formed, (d)-(e) the Hess
 diagrams of the observed and best-fit model CMDs, (f) the residuals, and
 (g) the significance of the residuals in Poisson $\sigma$. The cumulative
 mass fraction is shown in red in panel (a). The filled circles and error
 bars in panel (b) show the median metallicity and standard deviation in
 age bins representing at least 1~percent of the total mass of stars formed.}
\label{fig:a1}
\end{figure*}

\begin{figure*}
\includegraphics[width=15cm]{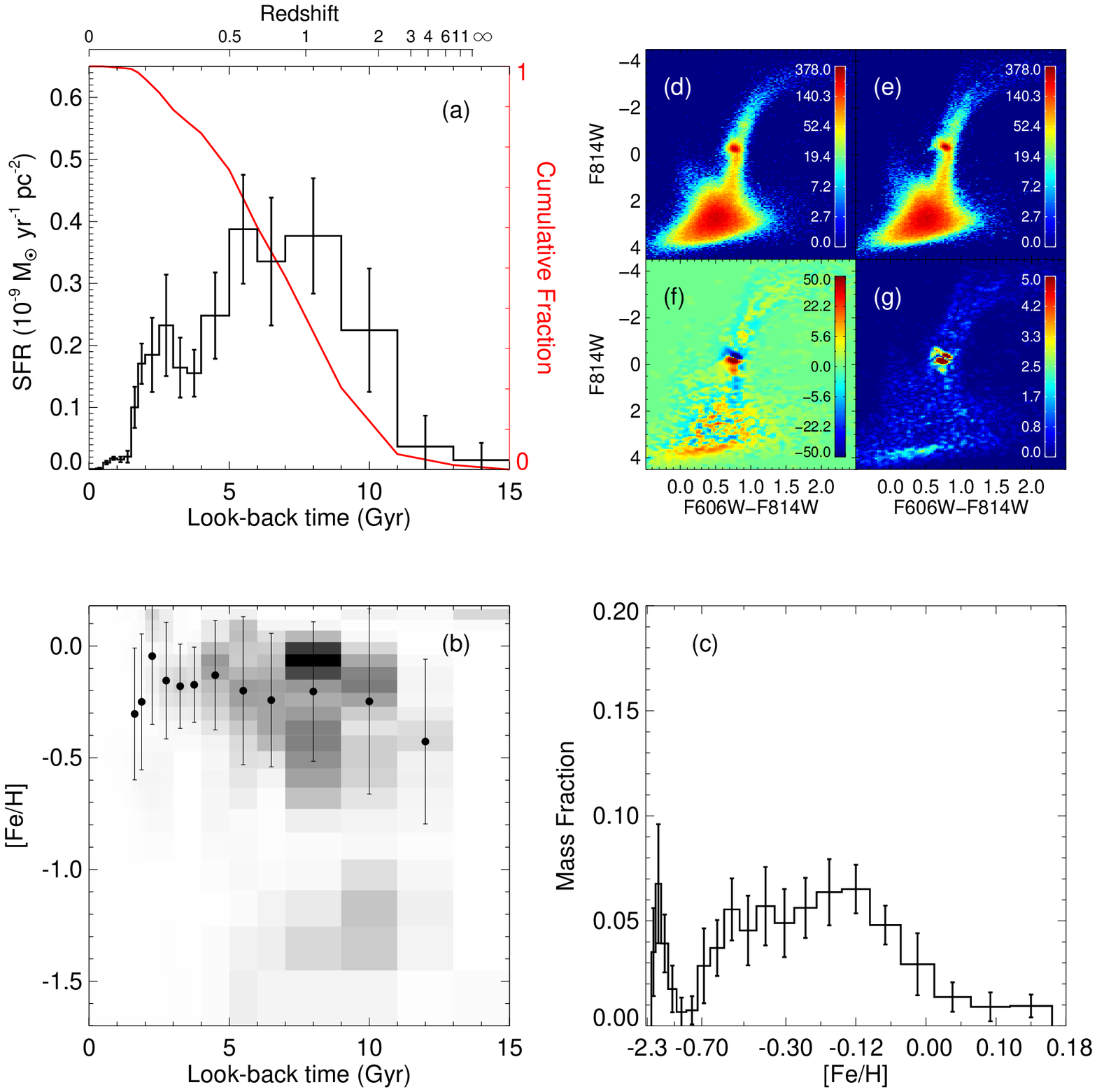}
\caption{Same as Fig.~A\ref{fig:a1}, but for the {\it NSpur}.}
\label{fig:a1b}
\end{figure*}

\begin{figure*}
\includegraphics[width=15cm]{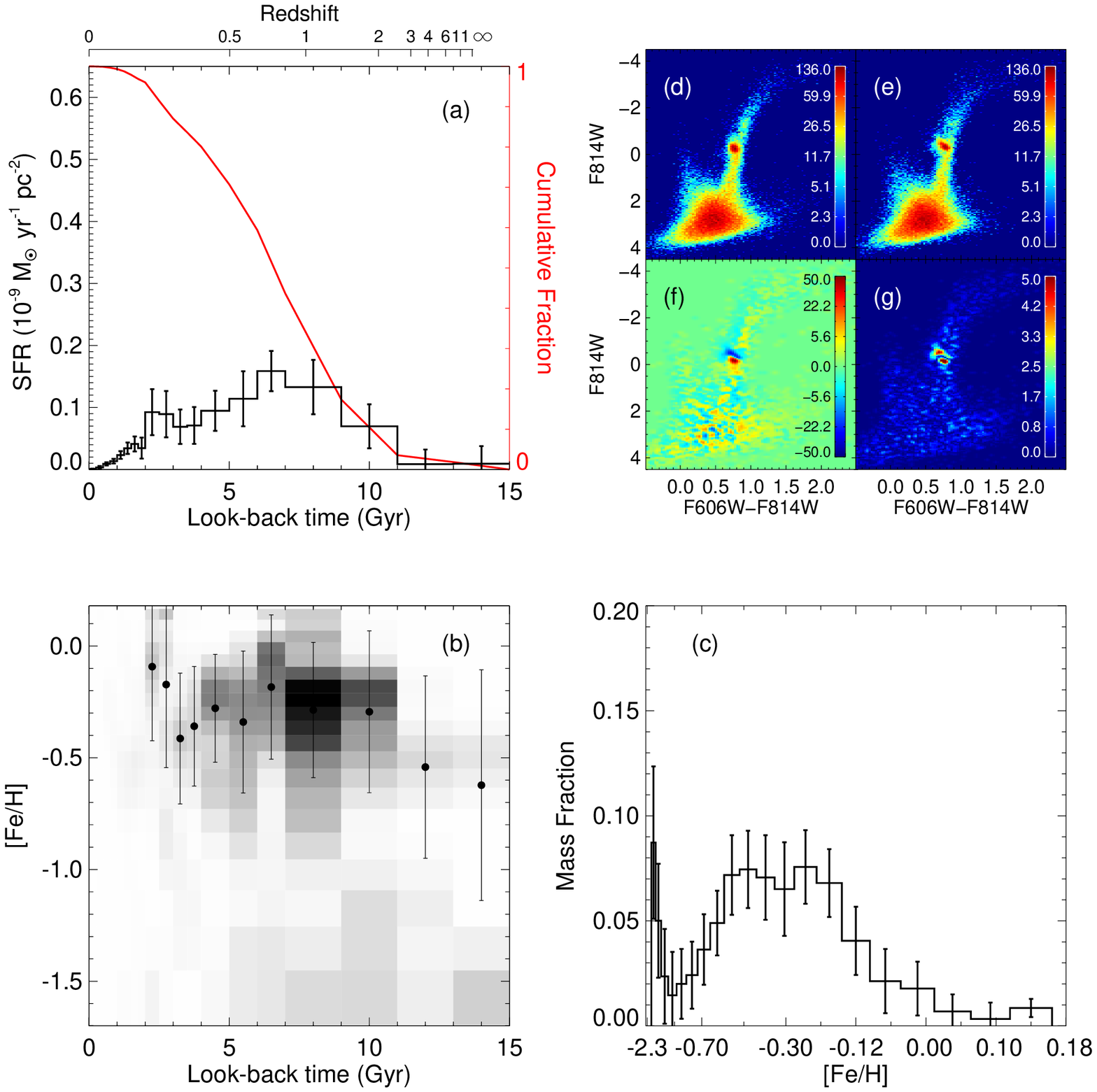}
\caption{Same as Fig.~A\ref{fig:a1}, but for the {\it G1 Clump}.}
\label{fig:a2}
\end{figure*}

\begin{figure*}
\includegraphics[width=15cm]{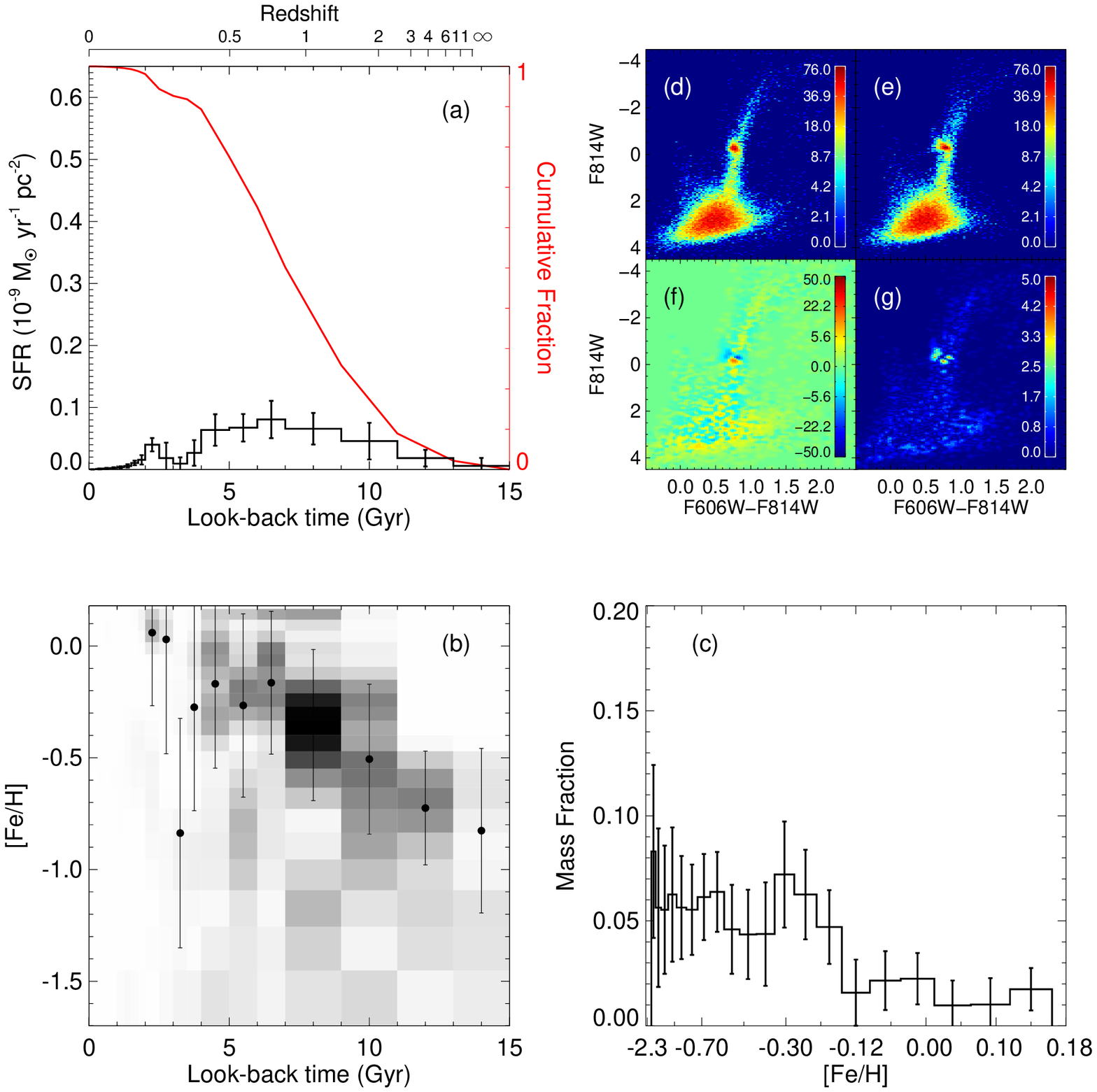}
\caption{Same as Fig.~A\ref{fig:a1}, but for the {\it NE Clump}.}
\label{fig:a2b}
\end{figure*}

\begin{figure*}
\includegraphics[width=15cm]{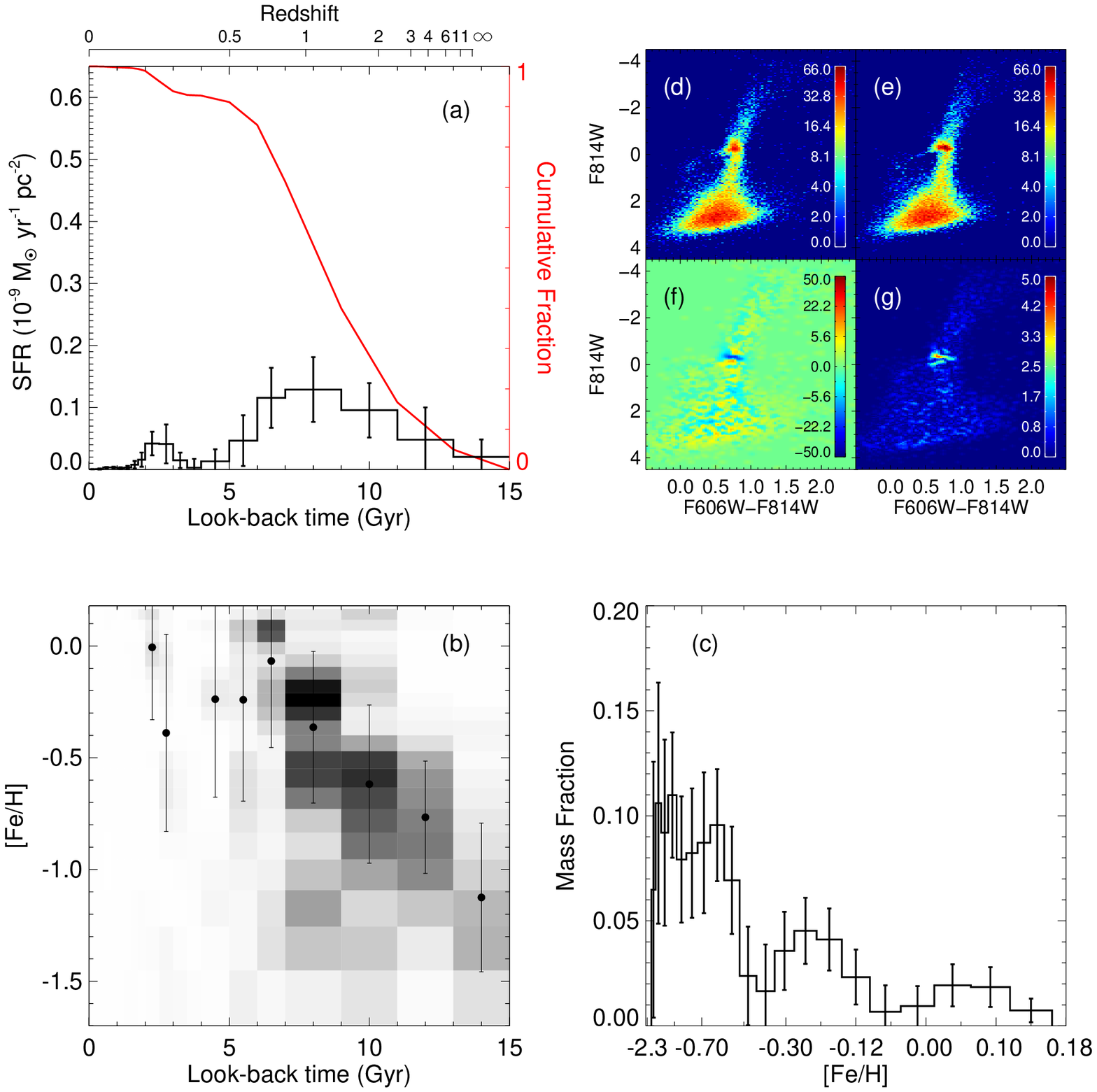}
\caption{Same as Fig.~A\ref{fig:a1}, but for the {\it EC1\_field}.}
\label{fig:a3}
\end{figure*}

\begin{figure*}
\includegraphics[width=15cm]{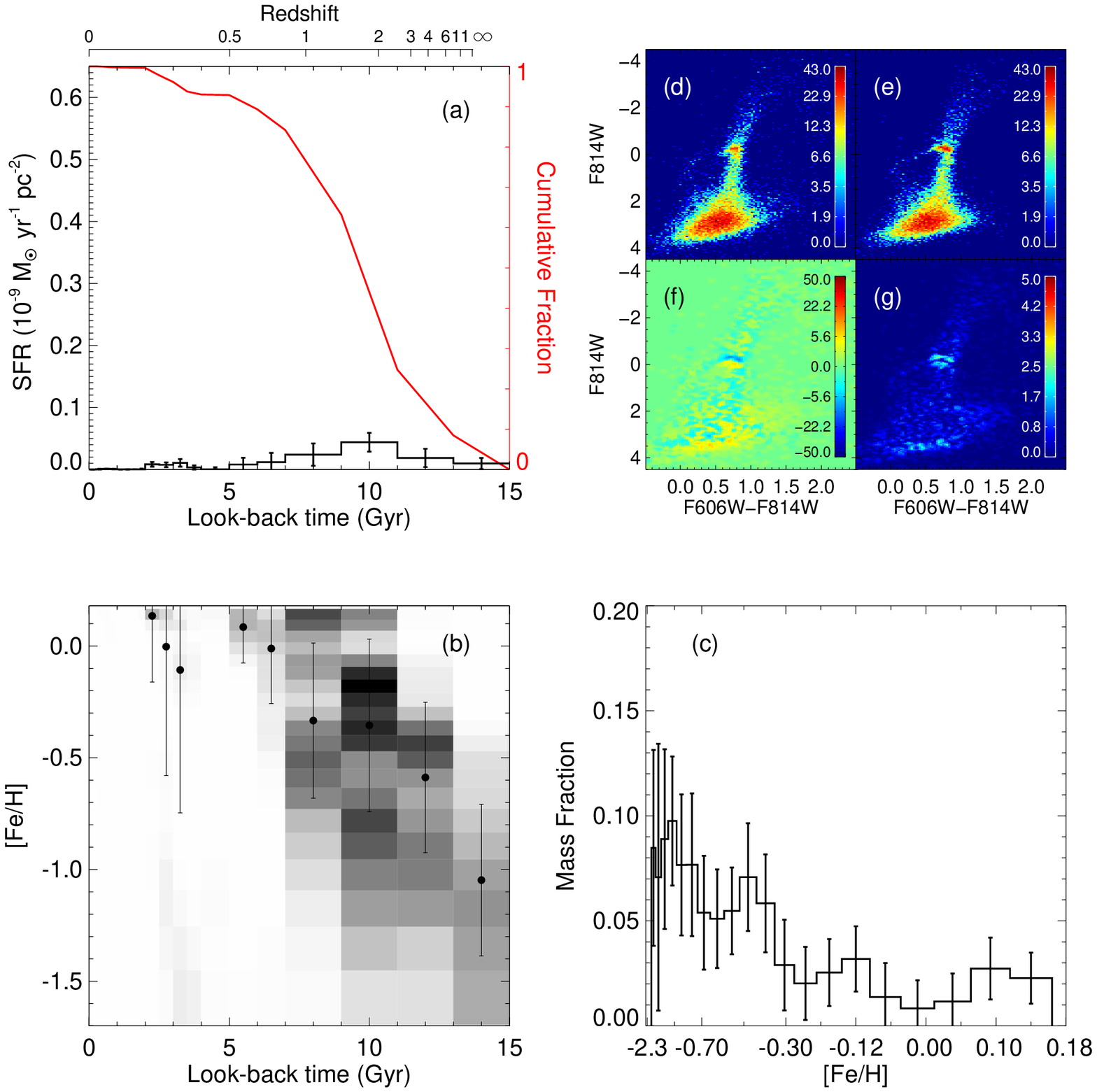}
\caption{Same as Fig.~A\ref{fig:a1}, but for the {\it Minor axis}.}
\label{fig:a3b}
\end{figure*}

\begin{figure*}
\includegraphics[width=15cm]{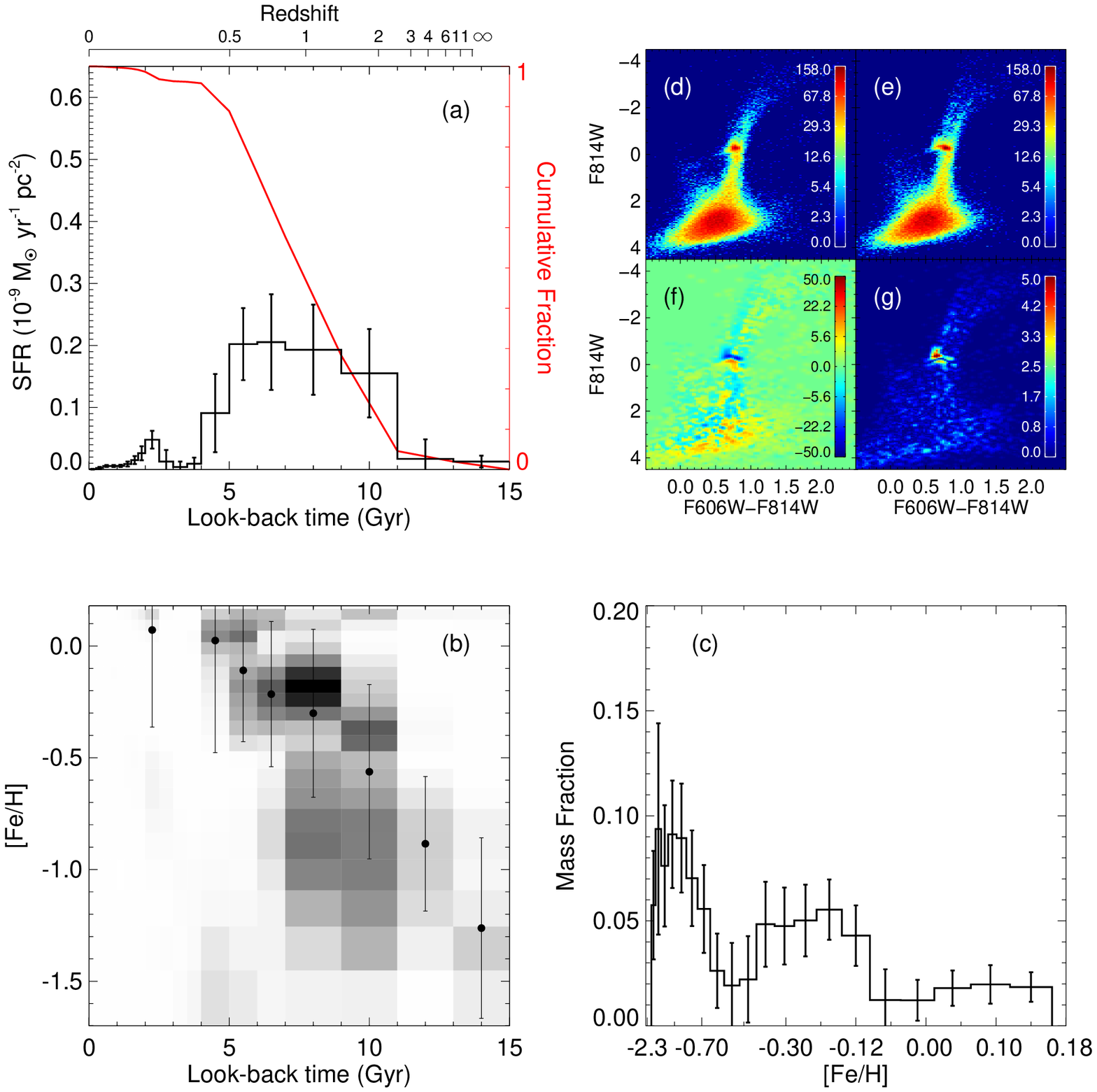}
\caption{Same as Fig.~A\ref{fig:a1}, but for the {\it NE Shelf}.}
\label{fig:a4}
\end{figure*}

\begin{figure*}
\includegraphics[width=15cm]{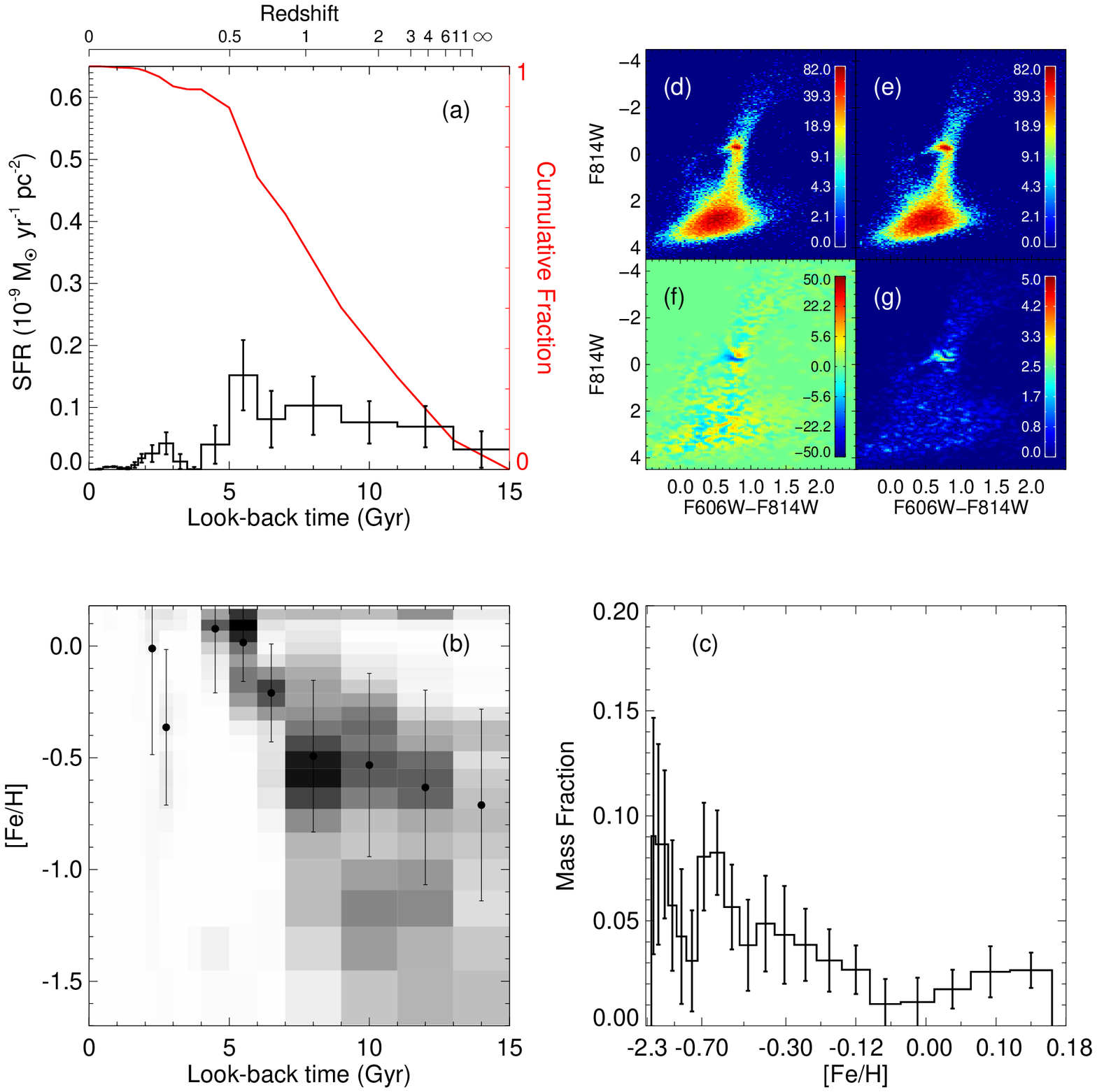}
\caption{Same as Fig.~A\ref{fig:a1}, but for the {\it Brown-stream}.}
\label{fig:a4b}
\end{figure*}

\begin{figure*}
\includegraphics[width=15cm]{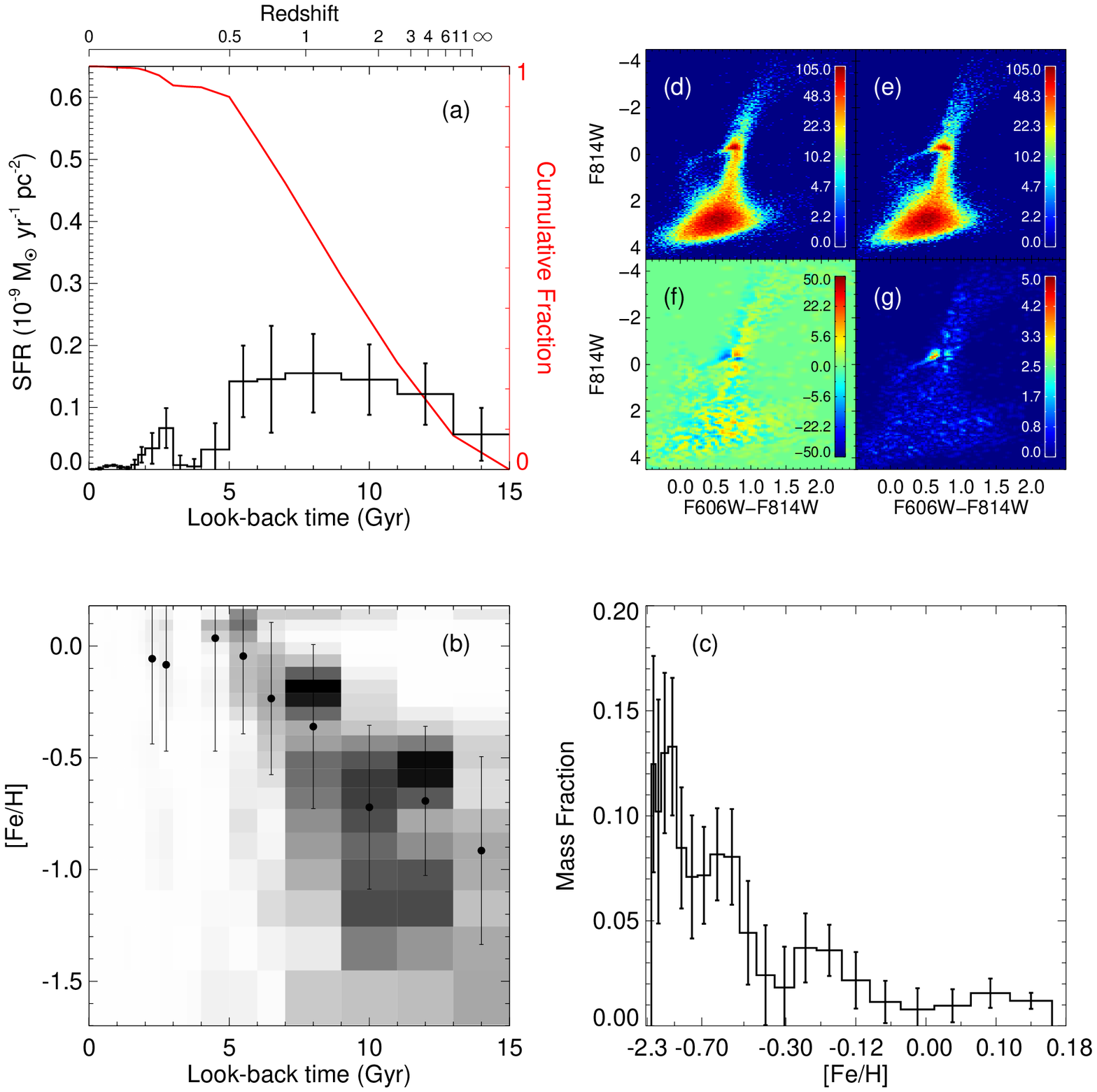}
\caption{Same as Fig.~A\ref{fig:a1}, but for the {\it Brown-spheroid}.}
\label{fig:a5}
\end{figure*}

\begin{figure*}
\includegraphics[width=15cm]{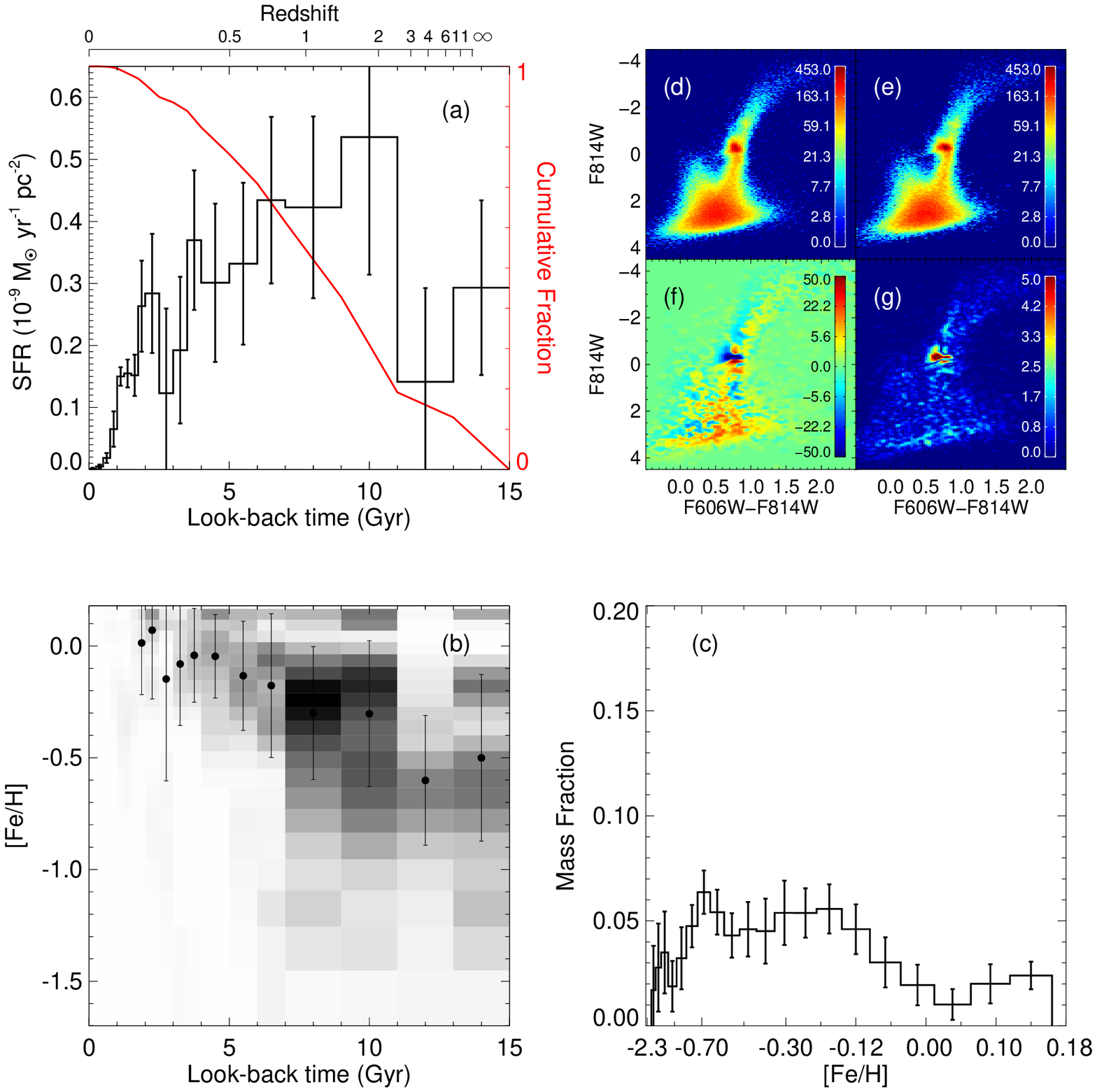}
\caption{Same as Fig.~A\ref{fig:a1}, but for the {\it GC6\_field}.}
\label{fig:a5b}
\end{figure*}

\begin{figure*}
\includegraphics[width=15cm]{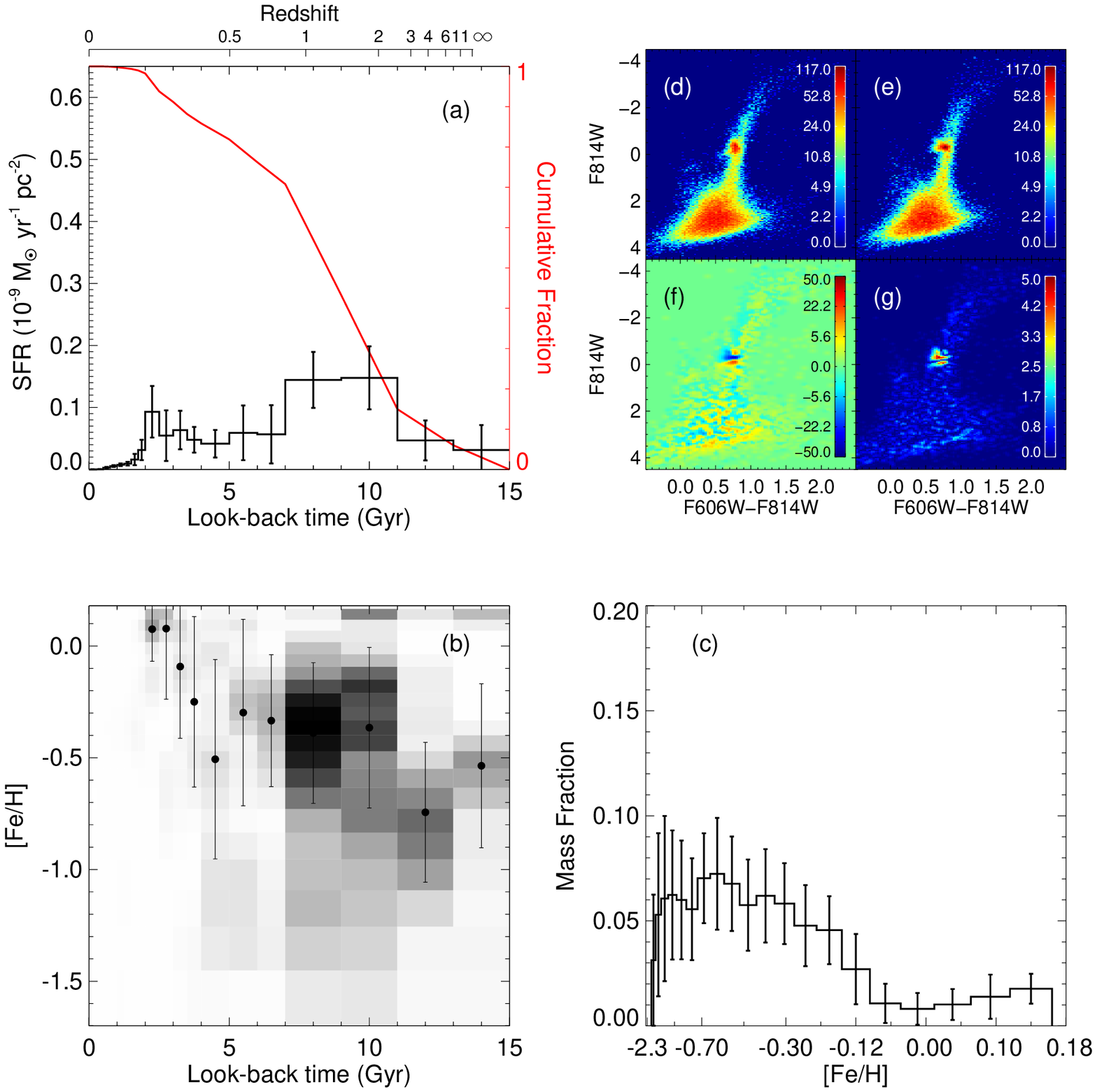}
\caption{Same as Fig.~A\ref{fig:a1}, but for the {\it NGC\,205 Loop}.}
\label{fig:a6}
\end{figure*}

\begin{figure*}
\includegraphics[width=15cm]{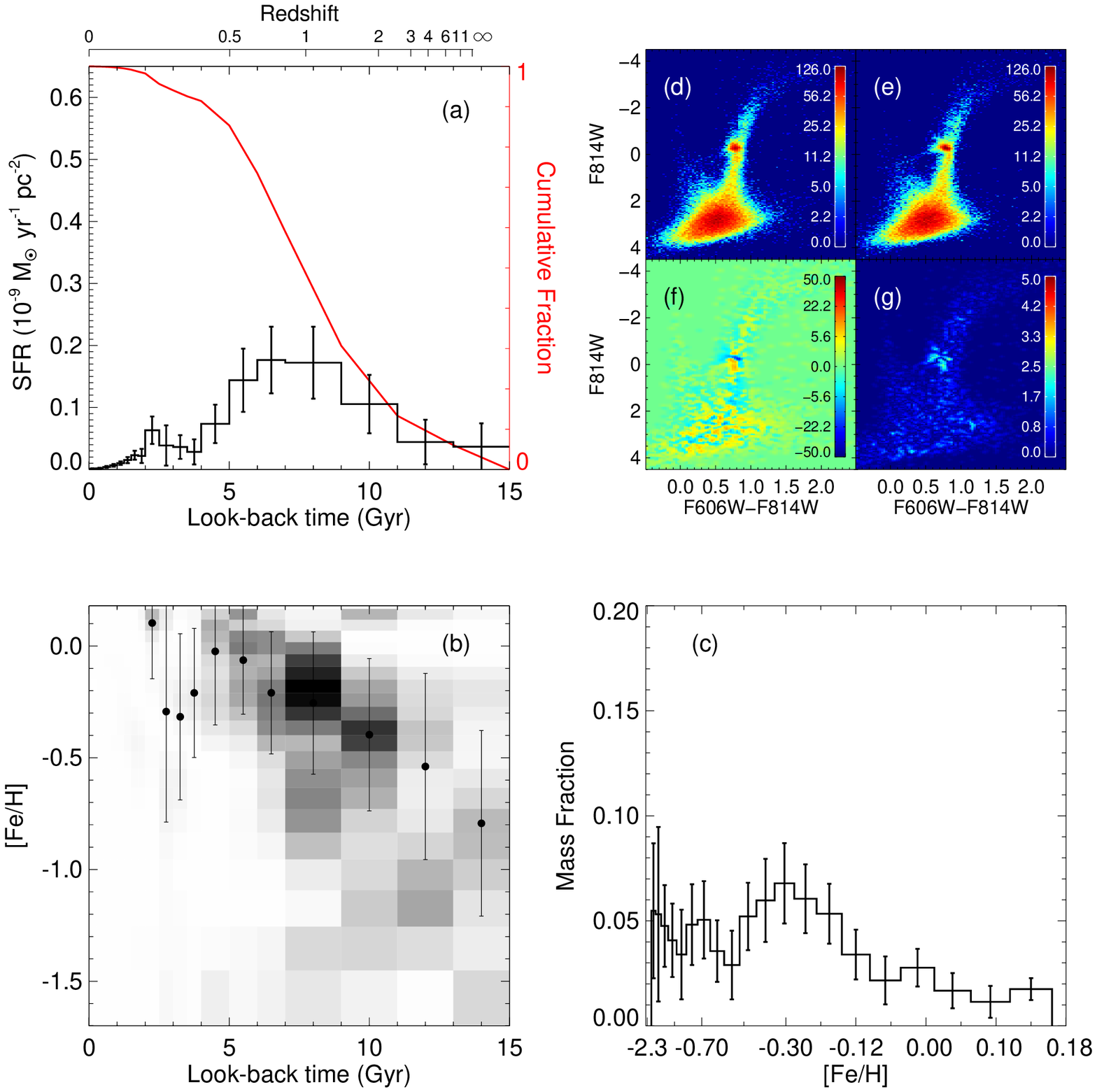}
\caption{Same as Fig.~A\ref{fig:a1}, but for the {\it Brown-disk}.
\label{lastpage}}
\label{fig:a6b}
\end{figure*}

\bsp

\end{document}